\documentclass[lettersize,journal]{IEEEtran}
\IEEEoverridecommandlockouts

\usepackage{graphicx} 
\usepackage{tabularx}
\usepackage{booktabs}
\usepackage{adjustbox}
\usepackage{framed}
\usepackage[table]{xcolor}
\usepackage{makecell}
\usepackage{url}
\usepackage{multicol}
\usepackage{multirow}
\usepackage{colortbl} 
\usepackage{tabularx}
\usepackage{newclude}
\usepackage{amsmath}
\usepackage{dingbat}
\usepackage[framemethod=TikZ]{mdframed}
\usepackage{enumitem}

\usepackage{lipsum}
\makeatletter  % Edit the following values as appropriate
\def\@IEEEBIOphotowidth{1in}    % width of the biography photo area 
\def\@IEEEBIOphotodepth{1in}   % depth (height) of the biography photo area
\def\@IEEEBIOhangwidth{1.2in}    % width cleared for the biography photo area
\def\@IEEEBIOhangdepth{1.25in}    % depth cleared for the biography photo area 
\makeatother

\usepackage{soul}

\newenvironment{keyfinding}[1][]{
    \mdfsetup{
        skipabove=5pt, % pads top.
        linecolor=blue,
        roundcorner=2pt,
        innerleftmargin=0.3cm,innerrightmargin=0.4cm, 
        linewidth=0.5pt,
        footnoteinside=false,backgroundcolor=blue!5}
    \small
    \begin{mdframed}}
    {\end{mdframed}}

\newcommand{\revised}[3]{%
  \begingroup
  \def\reviewerColor{}%
  \ifnum#2=1\def\reviewerColor{black}\fi
  \ifnum#2=2\def\reviewerColor{black}\fi
  \ifnum#2=3\def\reviewerColor{black}\fi
  \textcolor{\reviewerColor}{#3}%
  \endgroup
}

\newcommand{\minor}[3]{%
  \begingroup
  \def\reviewerColor{}%
  \ifnum#2=1\def\reviewerColor{black}\fi
  \ifnum#2=2\def\reviewerColor{black}\fi
  \ifnum#2=3\def\reviewerColor{black}\fi
  \textcolor{\reviewerColor}{#3}%
  \endgroup
}

\newcommand{\tracked}[1]{%
  \begingroup
  \def\reviewerColor{}%
  \textcolor{black}{#1}%
  \endgroup
}

\title{The Impact of Prompt Programming on Function-Level Code Generation}

\author{
Ranim Khojah$^1$, Francisco Gomes de Oliveira Neto$^1$, Mazen Mohamad$^{1,2}$, Philipp Leitner$^1$\\
$^1$\textit{Chalmers University of Technology and University of Gothenburg}, $^2$\textit{RISE Research Institutes of Sweden}\\
Gothenburg, Sweden \\
khojah@chalmers.se, francisco.gomes@cse.gu.se, mazen.mohamad@ri.se, philipp.leitner@chalmers.se}

\begin{document}

\maketitle

\begin{abstract}

% goal
% In this study, we aim to understand how different prompting techniques (and combinations of them) in a prompt impact the correctness, quality and similarity of the generated code.

Large Language Models (LLMs) are increasingly used by software engineers for code generation. However, limitations of LLMs such as irrelevant or incorrect code have highlighted the need for prompt programming (or prompt engineering) where engineers apply specific prompt techniques (e.g., chain-of-thought or input-output examples) to improve the generated code. 
% Despite this, the impact of different prompt techniques --- and their combinations --- on code generation remains underexplored.
\minor{R2C1}{2}{While some prompt techniques have been studied, the impact of different techniques --- and their interactions --- on code generation is still not fully understood.}
In this study, we introduce CodePromptEval, a dataset of 7072 prompts designed to evaluate five prompt techniques (few-shot, persona, chain-of-thought, function signature, list of packages) and their effect on the correctness, similarity, and quality of complete functions generated by three LLMs (GPT-4o, Llama3, and Mistral). Our findings show that while certain prompt techniques significantly influence the generated code, combining multiple techniques does not necessarily improve the outcome. Additionally, we observed a trade-off between correctness and quality when using prompt techniques. Our dataset and replication package enable future research on improving LLM-generated code and evaluating new prompt techniques.

\end{abstract}


\section{Introduction}

% LLMs in SE
With the widespread adoption of Large Language Models (LLMs) in software engineering, researchers and practitioners have uncovered their significant potential, particularly for code-related tasks, such as code generation and completion \cite{ross2023programmer, zeng2022extensive}. However, this adoption has also revealed several limitations of LLMs that can hinder developers' productivity \cite{khojah2024beyond} and cause frustrations \cite{weisz2022frustration}, preventing them from fully leveraging the benefits of LLMs in their coding process. Such limitations are related to hallucinations, misunderstanding the intent or purpose of the code, or simply generating incorrect code~\cite{zheng2023does}.

% prompt programming
These limitations are inherent to the design of LLMs, and are unlikely to "resolve themselves" entirely with future model generations. Therefore, researchers started proposing ways to mitigate these limitations by adapting how users interact with the LLMs. The interactions typically start with a natural language prompt that specifies what the LLM is expected to output. To ensure that LLM generates accurate, relevant, and high-quality outputs, users employ a structured approach to construct prompts, which is known as prompt programming.

% prompt techniques
To implement prompt programming, various prompt techniques can be used to guide the LLM on how to achieve the expected results \cite{white2023prompt, fiannaca2023programming, ahmed2024automatic}. For example, few-shot learning involves providing the LLM with a few input-output examples to guide the function logic, while adding context about the packages used can give the model additional information on what helper functions to use. 

% problem statement / motivation
However, such prompt techniques were evaluated based on the output accuracy for natural language generation tasks \cite{reynolds2021prompt, ahmed2024automatic} and are not well-studied for code generation, more specifically, function synthesis (generating function-level code), which is one of the most common use cases among software engineers \cite{khojah2024beyond}. Furthermore, evaluating the accuracy of code generation is not sufficient, since other aspects of the code are important for software engineers, such as maintainability and adherence to best practices.
Prompt techniques can also be combined \cite{white2023prompt}, but \revised{R1C1}{1}{to the best of our knowledge, no work evaluates the impact of multiple \textit{interacting} prompt techniques in one prompt. For instance, whether applying a certain prompt technique can cancel out, hinder, or even enhance the impact of an existing prompt technique in the prompt.}

Therefore, in this study, we design a full factorial experiment on five common prompt techniques for function generation along with all the possible combinations of these prompts, which sums up to 32 unique combinations of prompt techniques. To perform a comprehensive evaluation of the impact of different prompt techniques on code generation, we construct our dataset CodePromptEval which consists of 221 code-generation prompts from CoderEval \cite{Yu2024codereval}, that we extend with 32 possible variations for each prompt (that is, combinations of prompt techniques). This results in a total of 7072 datapoints.
We use CodePromptEval to generate functions with three popular LLMs (GPT-4o, Llama3, and Mistral), then evaluate the generated functions based on correctness, as well as quality and similarity to ground truth (e.g., in terms of naming style and structure). Particularly, we investigate the following research questions.\\

% With the widespread adoption of chatbots and Large Language Models (LLMs), many studies have investigated the impact of prompts and conversation flow on the interaction with chatbots in different tasks.
% Chatbots fail to provide accurate responses mainly due to misunderstanding the context and hence the intention of the user \cite{zheng2023does}. Therefore, current research focuses on the utilization of prompt engineering techniques and prompt structure to cover certain aspects that are needed to improve the performance and understanding of chatbots in various tasks \cite{white2023chatgpt, fiannaca2023programming, ahmed2024automatic}.

% mostly in NLP-related tasks such as Question/Answering.
% Although many of these tasks overlap with software-related tasks, the effectiveness of many prompt programming techniques may be limited to natural language-related tasks and possibly exclude code generation tasks.
% In our study, we present CodePromptEval, a dataset that provides 221 code-generation prompts. Each prompt has 32 possible variations and each variation employs a unique combination of prompt programming techniques (henceforth, prompt techniques). We use this dataset to perform a comprehensive evaluation of the impact of different prompt techniques on code generation. 

\noindent \emph{\textbf{RQ 1}: How do different LLMs perform on CodePromptEval?}

Initially, we study the performance of different current-generation LLMs (GPT-4o, Llama3, and Mistral) on our CodePromptEval dataset. We particularly look at the correctness of LLM-generated code as measured using existing test cases in CoderEval benchmark as a ground truth. We observe that the performance of all three evaluated LLMs is comparable, with a difference of around 5 percentage points between the best model, GPT-4o, and the worst, Mistral.\\

\noindent \emph{\textbf{RQ 2}: To what extent do different prompting techniques (and combinations of them) impact the code generation of LLMs?}

We now turn to the central research question of this paper. Using a full factorial experiment design, we compare how different prompt techniques (e.g., few-shot, providing a persona, etc.) impact the generated code in three dimensions: correctness, similarity to ground truth, and code quality.\\

 % We look into how certain properties of the LLM-generated code are affected by different prompts that use different prompting strategies , or combinations of them (e.g., few-shot learning with persona). We focus on three main properties of the code, that is, the correctness of the code, its quality, and its similarity to the code written by human developers. 

% \vspace{0.10cm}
\begin{quote}
\emph{\textbf{RQ 2.1}: How do prompt techniques impact the correctness of the code?}
\vspace{0.20cm}

To evaluate correctness, we test the functions, then measure the Pass@k scores for each combination of prompt techniques. We also perform statistical tests to identify the (combinations of) prompt techniques that impact the test results. We found that including only a function signature or few-shot examples has a significant positive impact on correctness. We further observe that combining prompt techniques \textit{does not} lead to significantly better results.
    % We also looked at the most common errors that occur among functions for the different prompts. We found that the most occurring error types when using LLMs to generating code are AssertionErrors, TypeErrors, AttributeErrors, and ImportErrors.
\end{quote}

\vspace{0.20cm}
\begin{quote}
\emph{\textbf{RQ 2.2}: How do prompt techniques impact the similarity of the code to a human-written baseline?}
\vspace{0.20cm}

We also study how similar generated solutions are to the (human-written) baseline. We find that including a persona, chain-of-thought, or signature increases the overall similarity to the baseline for some LLMs, while few-shot reduces only the lexical similarity. Note that generating code that is similar to an ``expected'' solution may be good or bad depending on context --- on the one hand, code that is close to the baseline may be easy to fix even if it is not passing the test cases; on the other, ``different'' can be particularly valuable if the goal is to brainstorm approaches, e.g., if used in ``exploration mode''~\cite{barke2023copilot}.
    % Regardless of the correctness of the code, we see how similar the generated code is to the ground truth. This can potentially tell us if the developers would need to make many modifications on the generated code, but also how different prompts can yield alternative approaches to solve the same problem. 
\end{quote}

\vspace{0.20cm}
\begin{quote}
\emph{\textbf{RQ 2.3}: How do prompt techniques impact the quality of the code?}
\vspace{0.20cm}

    Finally, we study code quality as measured through the presence of code smells and the (cyclomatic and cognitive) complexity of the code. We find that including a signature or few-shot examples leads to functions with higher complexity and more code smells. Interestingly, adding a relevant persona (``as a software developer who follows best coding practices ...'') indeed has a small positive effect on the code quality, but at the expense of slightly lower correctness.
\end{quote}
\vspace{0.20cm}

Overall, we conclude that the impact of prompt programming techniques is not dramatic for, at the time of writing, current-generation models. Most combinations of prompt techniques do not lead to statistically significant improvements (nor regressions) in correctness, similarity or quality. Providing type information for the function that is to be generated, either explicitly through a signature, or implicitly via few-shot examples, has the most clear effect. Some prompt techniques have a positive impact on correctness, and others on quality. However, the obvious idea of combining them usually improves neither.

% \noindent\textit{RQ3: What error types emerge when generating code using prompts with different prompt techniques?}

% 

% We assess the code generated by prompts that use a combination of different prompting strategies for code generation by looking at three different properties of the code. We focus on the correctness of the code (its ability to pass the assigned test), the quality (based on the cyclomatic complexity, cognitive complexity, and code smells), and its similarity to the code written by human developers (using CodeBLEU).

\section{Related work}

% LLMs in software engineering
Existing research on LLMs in software engineering has shown the potential of LLMs to support software engineers in various tasks, including requirements elicitation, software testing and documentation \cite{ronanki2023investigating, wang2024testing}. However, the main focus is directed towards code-related tasks \cite{khojah2024beyond}. This is also reflected in the interest among software organizations that, at the time of writing, leverage LLMs mostly for code generation, code completion and code summarization \cite{li2024fms}.
However, the increased adoption of LLMs for code-related tasks has unveiled risks and limitations, such as hallucinations, inaccuracies, and potential vulnerabilities \cite{toth2024llms, liu2024your}. Researchers have proposed the concept of \textit{prompt programming} (or prompt engineering) in order to minimize the model's limitations and trigger the LLM to output a more desirable response by using prompt techniques and provide relevant contextual information \cite{sahoo2024systematic, white2023prompt}.

% prompt techniques in research
Therefore, a new line of research emerged focusing on finding prompt techniques that can improve the performance of LLMs in various tasks. White et al. \cite{white2023prompt} propose different prompt patterns and techniques depending on the software-related task. 
However, the impact of these techniques on the LLM output can be unstable and inconsistent. Wang et al \cite{wang2024advanced} shows that prompt techniques can be sensitive to the specific task as well as the LLM (e.g., GPT-3.5 vs. GPT-4o). Other studies also show that the few-shot prompt technique \cite{brown2020fewshot} is effective, especially with the right structure \cite{fiannaca2023programming}, type \cite{margatina2023active} or order \cite{lu2022fantastically} of the examples (shots). Reynolds and McDonell \cite{reynolds2021prompt} highlight how few-shot examples can hurt the performance of the model and limit its search for a plausible solution in translation tasks. Contrastingly, we found that few-shot significantly improved the performance of the LLMs suggesting that prompt techniques have varied impact depending on the task and the domain.

% prompt engineering for code-related tasks
For code-related tasks, prompt techniques were shown to have a positive effect on code generation in the domain of education~\cite{wang2024enhancing}. Furthermore, researchers proposed ways and contextual information as prompt techniques to apply to the prompt and enhance code-related tasks \cite{shrivastava2023repository, ahmed2024automatic}, e.g., incorporating dataflow information to improve code summarization \cite{ahmed2024automatic}. Other prompt techniques used by Dong et al \cite{dong2024collaboration} included self-collaboration, where the LLM is prompted several times to take different personas e.g., first as a requirements engineer, then a software developer, then a tester, and only then return a code that resulted from the ``collaboration'' among the three personas. 
\revised{R1C1}{1}{Fagadau et al. \cite{fagadau2024influence} examined how individual prompt features (active voice and edge cases) affect code generation. Their experiments, which combined different prompt features, showed that most had little impact on the resulting code.}
In our study, we focus on three common prompt techniques, namely, few-shot learning, chain-of-thought \cite{cot}, and persona \cite{tseng2024persona}, as well as propose two pieces of contextual information as additional prompt techniques that are easily accessed by the developers, i.e., the imported packages and the signature of the function. \revised{R1C1}{1}{In addition, we investigate not only the impact of individual prompt techniques but also their interaction effects on code generation, for example, whether the effect of combining few-shot and persona arises from their interaction or from one technique alone.}

\begin{figure*}[!ht]
    \centering
    \includegraphics[width=0.84\linewidth]{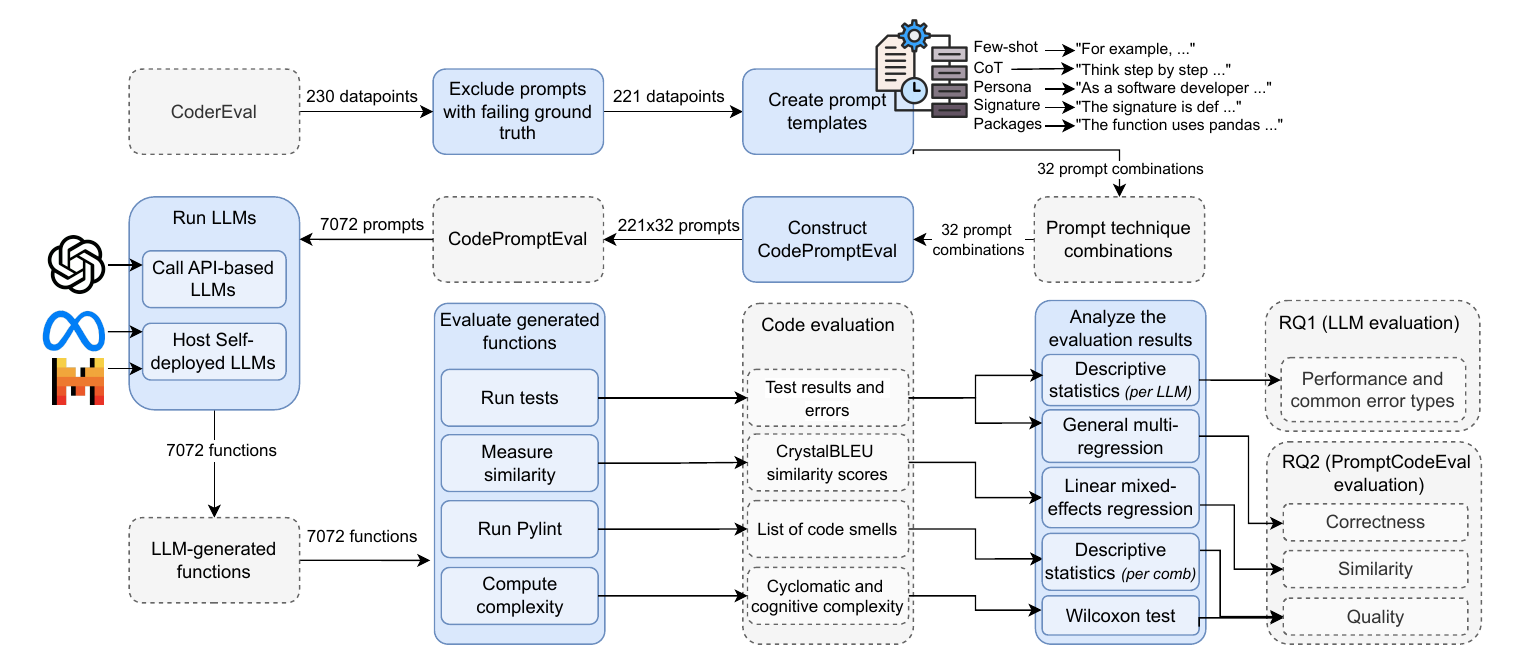}
    \caption{\protect\tracked{The process we follow to evaluate the code generated using different prompts by different LLMs (per run).}}
    \label{fig:process}
\end{figure*}

% evaluating code generation
To evaluate LLMs on code generation tasks, the most common metric is Pass@k~\cite{chen2021evaluating}, where k=1 is used to measure the rate of passed functions that the LLM generated on the first attempt~\cite{dongself2024passk2} (e.g., by running a test suite). CodeBLEU~\cite{ren2020codebleu} is another popular metric, commonly used in studies to measure the human-likenesses of generated code~\cite{wang2021codet5, jiang2024self}. Li et al.~\cite{li2024acecoder} conduct a manual human evaluation of their proposed prompt technique ``AceCoder'' based on correctness, presence of code smells, and maintainability.
We provide a systematic and automated approach to evaluate generated code based on correctness, maintainability, and similarity to the ground truth.

\section{Methodology}

Figure \ref{fig:process} shows our approach to evaluate the impact of commonly-used prompting techniques on the code generated by LLMs. On a high level, we create prompt templates that combine prompting techniques (e.g., CoT, few-shot, etc.) and apply each prompt template to 221 tasks from the CoderEval benchmark \cite{Yu2024codereval}. We evaluate three different LLMs (two open-weight and one proprietary), leading to 7072 generated functions per LLM. To understand the impact of each prompting technique and answer RQ2, we evaluate all functions in terms of correctness, similarity to the baseline of the benchmark, and quality using statistical analysis.

% \subsection{Dataset}
% In this experiment, we construct our CodePromptEval dataset which provides 32 variations\footnote{a variation is a prompt that applies a specific combination of prompt techniques. For example, ``Few-shot and Persona'' is a combination of prompt techniques and can be used to construct a prompt that applies these two techniques.} of each prompt in CoderEval -- an existing benchmark  \cite{Yu2024codereval}, then we run three decoder-only LLMs on CodePromptEval to generate a function per prompt variation. Finally, 
% % A fractional factorial experiment is a statistical experiment where only a subset of the factors are considered in a larger design space (i.e., the range of possible values of variables in the experiment)

We follow a full factorial experiment design and evaluate the code generation functionality of LLMs by varying two levels (present/absent) of five factors in a prompt, that is, the five prompt techniques: (1) few-shot learning, (2) Chain-of-Thought (CoT), (3) persona, (4) function signature, and (5) the list of packages. Therefore, we have 32 ($2^5$) treatments in our experiment. Note that the absence of all of these techniques counts as zero-shot, where only the generation instruction is present without any other prompt technique. We do not treat zero-shot as a factor since it cannot logically be combined with other prompt techniques (e.g., combining zero-shot with persona would simply default to persona). Instead, we use zero-shot as a baseline for prompt technique comparisons.

% \begin{table}[!ht]
% \caption{Scope and variables of our experimental study.}
% \label{tab:exp_comp}
% \begin{center}
% {
% \begin{tabularx}{\columnwidth}{p{2.7cm} p{5.3cm}}
% \toprule
% \textbf{Objective} & \textbf{Explore} \\
%  \midrule
%  Experimental Design: & Full Factorial Experiment \\
%  Experimental Units: &   Functions \\
% Experimental Subjects: &  CoderEval benchmark \\
% Dependent Variables:   &  Correctness, similarity, and quality.\\
% Factors: & Prompt programming technique \\
% Levels for Factors: & Few-shot, Chain-of-thought (CoT), Persona, Signature, Packages \\
% Parameters:  &  Programming language, transformer-based chatbots. \\
% \bottomrule
% \end{tabularx}
% }
% \end{center}
% \end{table}

% In this study, we aim to answer the following research questions:
% \begin{center}
% \textit{To what extent do different prompting techniques (or combinations of them) impact the code generation of chatbots?}
% % Does the performance of code generation increase with the increase of provided techniques/context?
% % What are the best combinations of prompt programming and context that are best for code generation?
% % Is prompt programming worth it?
% \end{center}

\subsection{Prompt technique combinations}
Prompt programming is the act of constructing a prompt using natural language to ensure that the model provides the intended response or to improve the performance of the model \cite{reynolds2021prompt}.
Based on observations from our previous work \cite{khojah2024beyond} and recommendations in literature \cite{white2023prompt} and from LLM providers such as OpenAI\footnote{\url{https://platform.openai.com/docs/guides/prompt-engineering}} and Microsoft\footnote{\url{https://learn.microsoft.com/en-us/azure/ai-services/openai/concepts/advanced-prompt-engineering}}\footnote{\url{https://microsoft.github.io/prompt-engineering/}} we decided on five prompt techniques to apply when prompting LLMs in our study. Examples for all prompt techniques will be provided later in Figure~\ref{fig:prompt-example}.

% In order to select the prompt programming techniques (prompt techniques for short), we look at recommendations from existing literature for various tasks, and at blogs about prompt programming by LLM providers such as OpenAI\footnote{\url{https://platform.openai.com/docs/guides/prompt-engineering}} and Microsoft\footnote{\url{https://learn.microsoft.com/en-us/azure/ai-services/openai/concepts/advanced-prompt-engineering}}\footnote{\url{https://microsoft.github.io/prompt-engineering/}}. 
% Based on observations from our previous work \cite{khojah2024beyond} and recommendations from LLM providers, we decided on five prompt techniques to apply when prompting LLMs in our study:

\begin{itemize}[leftmargin=8pt]
    \item \textit{Few-shot learning} 
    can be achieved by providing shots (or examples) to an LLM in order to enable learning new examples without the need to fine-tune the LLM \cite{brown2020fewshot}.
    \revised{R2C1}{2}{Typically, the examples describe the structure of the input and output. In code generation, such pattern will result in each example being composed of a generation task in natural language as an input, and a complete function as an output. We found this to be impractical from a user perspective, and poses a challenge of what generation tasks to choose in terms of the prompt design. Therefore, we follow an adapted pattern of few-shot prompting for code generation tasks also used in previous work~\cite{chen2021evaluating,xu2024fewshotcode}, where each example consists of a possible input of the function and its corresponding output.}
    We use two input-output examples explained in natural language. We do not consider a varying number of shots.
    \item \textit{Chain-of-Thought} (CoT) allows the LLM to break down the prompt by asking it \revised{R1C10}{1}{to ``think'' step by step before solving the problem. This technique is used to prompt the LLM to perform explicit reasoning~\cite{cot-org}.
    We apply Zero-shot-CoT \cite{cot} to isolate the impact of CoT from the few-shot prompt technique. In Zero-shot-CoT the steps that the LLM can follow are not explicitly mentioned in the prompt; rather, the model is expected to generate and follow its own reasoning process autonomously.}
    \item \textit{Persona} allows the LLM to play a specific role and consider its perspective when solving a problem \cite{wei2023leveraging, austin2021program}. For the persona, we use the role of a software developer who focuses on practices and standards that software developers follow.
    % \item \textit{Code purpose}: is a description in the natural language of the overall goal and purpose of the code \cite{bareiss2022description}.
    \item \textit{Signature} is a line of code that includes the signature of the function to generate. The signature includes the function name, the input parameters, and (optionally) the output.
    \item \textit{Packages} is a list of libraries and files that exist in the environment in which the code runs. This includes local packages and external libraries. \revised{R1C9}{1}{The packages used by a function are extracted by parsing all import statements in the Python file to collect the names of the modules and external libraries on which the function depends.}
    % \revised{R1C9}{1}{The packages used by a Python file are extracted by systematically parsing its import statements to collect the names of the modules and external libraries it depends on.}
\end{itemize}

\minor{R1C1}{1}{Previous work indicates that Signature and Packages are not necessarily used when prompting LLMs for code generation by developers \cite{khojah2024beyond}. While Signature is often a part of the prompt in code completion benchmarks \cite{athiwaratkun2022multihumaneval}, most code generation benchmarks construct prompts based on documentation (e.g., docstrings) without the signature ~\cite{Yu2024codereval}.
% Signatures and documentation (e.g., docstrings) have been used in HumanEval for code completion~\cite{athiwaratkun2022multihumaneval}, whereas CoderEval leverages code generation based on the functions' documentation instead of the signature ~\cite{Yu2024codereval}.
Therefore, in this study, we present them as prompt techniques for code generation tasks as they might provide additional context that can guide the model toward more relevant code generation (e.g., types inferred by the parameter names).}

% We also consider zero-shot as a baseline to the other techniques. In the scope of this study, zero-shot refers to the prompt with the code generation instruction, without applying any of the five prompt techniques above.

\subsection{CodePromptEval}
To evaluate the different combinations of prompt techniques, we construct CodePromptEval -- a dataset that includes 221 function-level code generation tasks, where each generation task is implemented using 32 different prompt variations. Each of these 32 prompts applies a unique combination of prompt techniques, resulting in a total of 7072 prompts (221 tasks $\times$ 32 variations).

% a dataset that includes 7072 prompts representing 221 prompts of distinct generation tasks along with 32 variations of each prompt. Each variation applies a unique combination of prompt techniques.

% eliminate datapoints
To create our dataset, we initially start with the CoderEval Python dataset \cite{Yu2024codereval}. This dataset consists of 230 datapoints from 43 Python projects. Each datapoint consists of a prompt, a Python function (human-written baseline), and the corresponding tests (in form of unit tests or a main class).
We first set up different virtual environments for functions from different projects, then we test the functions using the provided tests, and eliminate nine datapoints where the baseline does not pass the tests. This resulted in 221 datapoints that will be the foundation for our own CodePromptEval dataset.

% clean prompts
Then, we ensure that the prompts are ``pure'' from any prompt technique that may be implicitly applied (e.g., providing examples), by going through the prompts manually and removing any elements that do not describe the purpose of the code. We then treat this prompt as a zero-shot prompt.

% extract info and map it to prompt techniques.
The next step was to prepare prompt templates by defining how each prompt technique will be implemented and mapping relevant information to prompt techniques.
In particular, for each datapoint, we extract the signature of the function and the list of used packages (represented as imports at the beginning of the class). For chain-of-thought, we adapted the template recommended by Zhuosheng et al. \cite{cot}. To construct the persona, we defined a persona description of a software developer who follows best coding practices
for maintainability. 
% create few-shot examples
To implement the few-shot prompt technique, the first three authors of this paper manually constructed two input-output examples for each prompt following the template ``If the input is X, then the output is Y''. We also create corresponding tests to ensure that the input and output are correct. \revised{R1C10}{1}{The examples were created based on the goal of covering both a typical (mainline) case and an edge case to ensure that the examples capture a range of expected behavior.}

\begin{table}[ht!]
\scriptsize
\begin{center}
\caption{The 32 combinations of prompt techniques that we consider in our full factorial experiment.}
\label{tab:combinations} 
\begin{tabular}{c|ccccc}
\toprule
\textbf{ID} & \textbf{Few-shot} & \textbf{CoT} & \textbf{Persona} & \textbf{Packages} & \textbf{Signature} \\
\cmidrule(lr){1-1} \cmidrule(lr){2-6} 
\textbf{P1} & - & - & \checkmark & \checkmark & \checkmark \\
\textbf{P2} & - & - & \checkmark & \checkmark & - \\
\textbf{P3} & - & - & \checkmark & - & \checkmark \\
\textbf{P4} & - & - & \checkmark & - & - \\
\textbf{P5} & - & - & - & \checkmark & \checkmark \\
\textbf{P6} & - & - & - & \checkmark & - \\
\textbf{P7} & - & - & - & - & \checkmark \\
\textbf{P8} & - & - & - & - & - \\
\textbf{P9} &- & \checkmark & \checkmark & \checkmark & \checkmark \\
\textbf{P10} & - & \checkmark & \checkmark & \checkmark & - \\
\textbf{P11} & - & \checkmark & \checkmark & - & \checkmark \\
\textbf{P12} & - & \checkmark & \checkmark & - & - \\
\textbf{P13} & - & \checkmark & - & \checkmark & \checkmark \\
\textbf{P14} & - & \checkmark & - & \checkmark & - \\
\textbf{P15} & - & \checkmark & - & - & \checkmark \\
\textbf{P16} & - & \checkmark & - & - & - \\
\textbf{P17} & \checkmark & - & \checkmark & \checkmark & \checkmark \\
\textbf{P18} & \checkmark & - & \checkmark & \checkmark & - \\
\textbf{P19} & \checkmark & - & \checkmark & - & \checkmark \\
\textbf{P20} & \checkmark & - & \checkmark & - & - \\
\textbf{P21} & \checkmark & - & - & \checkmark & \checkmark \\
\textbf{P22} & \checkmark & - & - & \checkmark & - \\
\textbf{P23} & \checkmark & - & - & - & \checkmark \\
\textbf{P24} & \checkmark & - & - & - & - \\
\textbf{P25} & \checkmark & \checkmark & \checkmark & \checkmark & \checkmark \\
\textbf{P26} & \checkmark & \checkmark & \checkmark & \checkmark & - \\
\textbf{P27} & \checkmark & \checkmark & \checkmark & - & \checkmark \\
\textbf{P28} & \checkmark & \checkmark & \checkmark & - & - \\
\textbf{P29} & \checkmark & \checkmark & - & \checkmark & \checkmark \\
\textbf{P30} & \checkmark & \checkmark & - & \checkmark & - \\
\textbf{P31} & \checkmark & \checkmark & - & - & \checkmark \\
\textbf{P32} & \checkmark & \checkmark & - & - & - \\
\bottomrule
\end{tabular}
\end{center}
\vspace{-0.7cm}
\end{table}

% create combinations
Finally, we define 32 prompt variations that we list in Table \ref{tab:combinations}. Each variation represents a prompt that applies a unique combination of prompt techniques. For example, \textbf{P7} is a prompt that provides the code signature, but uses no other prompt programming technique, whereas \textbf{P28} combines few-shot learning with CoT and the usage of the persona ``software developer''. \textbf{P8} is the zero-shot baseline, where no prompt technique is used and the model is only provided with the programming task. \textbf{P25} is the case where \emph{all} prompt techniques are used in conjunction.

% constructing CodePromptEval
We then map each variation from Table~\ref{tab:combinations} to the relevant information and templates for prompt techniques (e.g., imported libraries for packages), then we combine them with the 221 prompts from CoderEval, \revised{R1C4}{1}{leading to CodePromptEval with 7072 concrete prompts (221 prompts times 32 variations) and their corresponding Python functions as ground truth}. 

\revised{R1C4}{1}{The functions include both domain-specific implementations and commonly used utility logic extracted from GitHub repositories, rather than standard textbook algorithms such as sorting or searching. Table \ref{tab:codereval-stat} describes the functions' length and cyclomatic complexity.
Moreover, the 221 functions fall into six code dependency levels: 33 self-contained (does not use packages outside the function scope), 25 standard library runnable (uses libraries available as part of Python standard library), 19 public library runnable (uses libraries available on PyPI), 54 class runnable (uses code outside the function, but within the class), 67 file runnable (uses code outside the class, but within the file), and 23 project runnable (uses code in other files).}
Our dataset, the virtual environments, and the few-shot examples and tests are provided in our replication package~\cite{khojah2024replication}.

\begin{table}[h]
\scriptsize
\centering
\caption{
\protect\revised{R1C4}{1}{Function Statistics in CodePromptEval}
}
\begin{tabularx}{\linewidth}{lXXXX}
\toprule
\textbf{\tracked{Metric}} & \textbf{\tracked{Min}} & \textbf{\tracked{Max}} & \textbf{\tracked{Mean}} & \textbf{\tracked{Std Dev}} \\
\midrule
\tracked{Number of Variables} & \tracked{0} & \tracked{32} & \tracked{2.52} & \tracked{4.35} \\
\tracked{Number of Parameters} & \tracked{0} & \tracked{7} & \tracked{1.70} & \tracked{1.30} \\
\tracked{Lines of Code} & \tracked{3} & \tracked{564} & \tracked{32.14} & \tracked{55.82} \\
\tracked{Block Depth} & \tracked{1} & \tracked{9} & \tracked{2.58} & \tracked{1.63} \\
\tracked{Cyclomatic Complexity} & \tracked{1} & \tracked{29} & \tracked{4.57} & \tracked{4.66} \\
\bottomrule
\end{tabularx}
\label{tab:codereval-stat}
\end{table}

% 230 functions from 43 Python projects and 230 methods from 10 Java projects.
% To create our dataset, we first extract different information from the prompt description and/or the ground truth code of  CoderEval \cite{Yu2024codereval}, a Python-based benchmark for code generation containing 230 datapoints (generation problems). In particular, we extract for each datapoint (1) the programming language, (2) used packages, (3) the original docstring, (4) the signature of the function, (5) the ground truth implementation, (6) examples of possible input/output (if existing for this datapoint in CoderEval), and (7) the corresponding test code (also if existing for this datapoint in CoderEval).

% We construct CodePromptEval by first generating each combination of the prompt programming techniques listed in Section~\ref{sec:techniques}. These 32 variations are listed in Table \ref{tab:combinations}. For example, \textbf{P7} is a prompt that provides the code signature, but uses no other prompt programming technique, whereas \textbf{P28} combines few-shot learning with CoT and the usage of the persona "software developer".
% We then combine each variation from Table~\ref{tab:combinations} with each datapoint from CoderEval, leading to 7072 concrete prompts (221 datapoints times 32 variations).

\begin{figure}[!ht]
\vspace{-0.6cm}
    \centering
    \includegraphics[width=\linewidth]{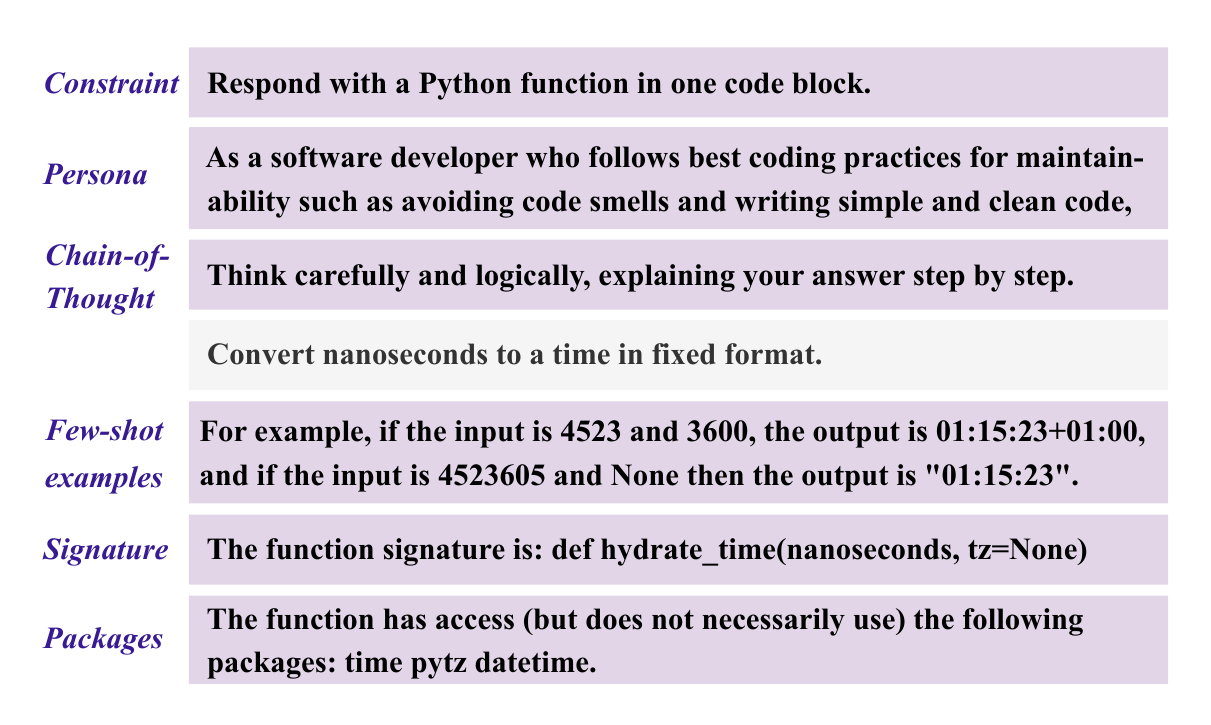}
    \caption{Example prompt in CodePromptEval.}
    \label{fig:prompt-example}
\end{figure}

We illustrate an example prompt with all prompting techniques (\textbf{P25}) in Figure \ref{fig:prompt-example}. 
% Purple text in square brackets (e.g., [Persona]) is provided for illustration purposes and not part of the actual prompt that will be used in the experiment. 
The prompt description, signature, and packages are extracted from CoderEval, while we construct the few-shot examples, persona, and chain-of-thought texts as a part of CodePromptEval.
We also append a constraint at the beginning of each prompt to ensure that the output has a block of Python code with a self-contained function. 
% This constraint is used in all combinations.

If different prompt techniques are combined, we apply them in a fixed order (as given in Figure \ref{fig:prompt-example}). This order ensures the sentence flows naturally. For example, the common practice is to place the persona at the beginning, and the few-shot examples after the purpose of the code. While it is possible to experiment with different orders of prompt techniques in a prompt, we consider this outside the scope of this study.

\subsection{Code Generation}
We focus on LLMs with a decoder-only transformer architecture, which is at the time of writing the preferred architecture to use in code generation tasks \cite{jiang2024survey}. Therefore, we select the following LLMs for our study: GPT-4o, Llama3-70B-Instruct, and Mistral-Small-Instruct-2409 (22B). We also collect data for two previous-generation LLMs (GPT-3.5-turbo and Llama2-7B-Instruct), but omit discussing the results for these older models for reasons of brevity in this paper.
\revised{R1C11}{1}{In general, the results for these older models showed lower passing rates and higher complexities. However, the overall impact of the techniques remained consistent with the findings we report for the studied LLMs, albeit with varying significance levels.}
The collected data for these models is still available in our replication package~\cite{khojah2024replication}.

% In this paper, we present the results for GPT-4o, Llama3-70B-Instruct and Mistral. While we include the results of the older generation LLMs GPT-3.5-turbo and Llama2-7B-Instruct in our replication package.

\tracked{We run all 7072 prompts on the selected LLMs three times to account for the non-deterministic nature of LLMs. For all LLMs, we set the temperature to 0.2, which has been commonly used for code generation tasks \cite{chen2021evaluating, temp1}, and complies with the recommendations for our correctness measure~\cite{chen2021evaluating}}. For the API-based GPT models, we send requests to the external API and store the responses. We host the remaining models on the Alvis cluster, a NAISS resource (National Academic Infrastructure for Supercomputing in Sweden) dedicated to Artificial Intelligence and Machine Learning research\footnote{\url{https://www.c3se.chalmers.se/about/Alvis/}} using models downloaded from Huggingface\footnote{\url{https://huggingface.co}}. Running the self-hosted LLMs on Alvis required around \tracked{2800} GPU hours using Nvidia A100 GPUs. For the GPT models, we use the OpenAI API, which is billed based on the tokens that are processed. To run GPT-4o and GPT-3-turbo on all the prompts in CodePromptEval, we provide around \tracked{2.45} million input tokens and generate approximately \tracked{7.66} million output tokens.
% 1,820 GPU hours
% Input tokens: 1,100,430
% Output tokens: 3,732,503

\subsection{Evaluating the LLM-generated functions}
After generating 7072 code solutions \tracked{three times}, we evaluate them based on three main aspects following our research questions (correctness, similarity, and quality). 
We use different tests to measure statistically significant differences for the measures below, hence we detail the choice of statistical methods in their corresponding results sections.\\

% \subsubsection*{Code characteristics}
% We calculate the length of the code in terms of tokens as well as label the generated function with the code level that maps to the ground truth code level that we obtain from the CoderEval dataset.

\subsubsection*{Correctness}

To evaluate their correctness, we run the generated functions against their corresponding tests in CoderEval. \revised{R1C2}{1}{When running the functions, we replace the generated function name with the originally expected one to ensure compliance with our test cases.}

There are two types of tests in CoderEval: Python unit tests, and a main function with different statements and conditions that set a boolean variable \texttt{isT} (is True) to \texttt{False} when at least one of the conditions does not hold. To ensure consistency and instrumentation of our experiment, ensure that an AssertionError is thrown when needed, by adding an assert statement at the end of tests in the form of a main function \texttt{assert isT}. Furthermore, as some of the LLM-generated functions can be erroneous and get stuck in an infinite loop, we wrap the tests with error-handling constructs (a \texttt{try/except}) and set a timeout of 60 seconds per function. Then we collect the test results and the error messages when applicable.

\revised{R1C5}{1}{We distinguish syntactic correctness and semantic correctness of the function. The Python function is syntactically correct if its syntax is valid and the function is runnable. Semantically correct functions are functions that pass their corresponding tests. When a function is both syntactically and semantically correct, then it is labeled as plausibly correct \cite{corso2024correctness}. In the remainder of the paper, we use correctness as a short-hand for plausibly correct.}

\subsubsection*{Similarity}
\revised{R1C6}{1}{To assess the LLM-generated code's similarity to the ground truth obtained from CoderEval (human-written functions), we measure the CrystalBLEU score~\cite{crystalbleu}. CrystalBLEU combines four n-gram measures where $n=1,2,3,4$ while accounting for ``trivial grams'' that are shared across all functions, such as Python keywords. The combined n-grams that are used as a metric for syntactic similarity are then used as a proxy to estimate the semantic similarity.}

\subsubsection*{Quality}
Regarding code quality, we focus on measures that are related to maintainability~\cite{heitlager2007practical} and we only measure them for functions that pass their tests. In other words, we measure the quality only for functionally correct functions. We use Pylint\footnote{\url{https://pypi.org/project/pylint/}} to generate a report with identified code smells in the generated functions.
Moreover, we compute the code complexity for both the LLM-generated functions and the equivalent ground truth (i.e., the human-written functions in CoderEval) to compare both results and see how the different prompts have an impact on the code quality. 
Code complexity refers to how detailed and interconnected different parts of the code are, which can make the code harder to understand and test. To get an overview of the complexity of the generated functions, we measure McCabe's cyclomatic complexity via the Radon Python package\footnote{\url{https://pypi.org/project/radon/}} and cognitive complexity~\cite{campbell2018cognitive} via the cognitive-complexity Python package.\footnote{\url{https://pypi.org/project/cognitive-complexity/}}

\section{CodePromptEval Overview}

% In this section, we present the evaluation of the LLMs' performance on CodePromptEval in Section \ref{sec:llm-eval}, then for each LLM, we present the evaluation results for different combinations of prompts in CodePromptEval in terms of correctness, similarity, and quality in Section \ref{sec:codepromteval-eval}. Finally, we provide an extended analysis of the different error types triggered by CodePromptEval in Section \ref{sec:error-eval}.

In this section, we provide an overview of the aggregate results from running three LLMs (GPT-4o, Llama-3, and Mistral) \tracked{three times} on the CodePromptEval dataset. This section answers RQ1 in our study.

Note that these results are not an assessment of the capabilities of these models when used with an ``ideal'' prompt, but an aggregation over all prompt technique combinations in our study. That is, the following results should be read as an overview of CodePromptEval, and not as a judgment of which LLM performs best in general.
Detailed drill-downs assessing the performance of individual (combinations of) prompt techniques will be presented in Section~\ref{sec:codepromteval-eval}.

% \subsection{Evaluation of the LLMs}
% \label{sec:llm-eval}
% We present the test results obtained from running GPT-4o, Llama-3, and Mistral on CodePromptEval . The results in this section are aggregated results for all prompts and programming techniques. A more fine-grained discussion of differences between individual (combinations of) techniques will be presented in the following sections.

% \begin{table}[!ht]

%     \begin{center}
%     \caption{Overview of passed and failed functions per LLM. The total number of functions per LLM is 7072.}
%     \label{tab:test-res-llms}
%     \begin{tabularx}{\columnwidth}{lrr}
%     \toprule
%          \textbf{LLM}                &  \textbf{\# Passed Functions}     & \textbf{\# Failed Functions}  \\
%     \midrule
%          \textbf{GPT-4o}             &   3707  (52.42\%)       & 3365 (47.58\%)      \\
%          % GPT-3.5-turbo      &   3708 (52.43\%)        & 3364 (47.57\%)      & 7072 \\
%          \textbf{Llama3-70B-Instruct} &  3564 (50.40\%)        & 3508 (49.60\%)      \\
%          % Llama2-7B-Instruct &   2971 (42.01\%)        & 4101 (57.99\%)      & 7072 \\
%          \textbf{Mistral-22B-Instruct} &  3335 (47.16\%)        & 3737 (52.84\%)      \\
%     \bottomrule
%     \end{tabularx}

%     \end{center}
% \end{table}

\begin{table}[!ht]
\scriptsize
    \begin{center}
    \caption{\protect\revised{R1C5}{1}{Overview of passed and failed functions per LLM. We also show the breakdown of failures types (total = 7072 functions, averaged over 3 runs)}}
    \label{tab:test-res-llms}
    \begin{tabularx}{\columnwidth}{lrrr}
\toprule
\textbf{Results} & \textbf{GPT-4o} & \textbf{Llama3-70B} & \textbf{Mistral-22B} \\
\midrule
\textbf{Passed} & 3691 $\pm$ 12 (52.2\%) & 3575 $\pm$ 11 (50.5\%)  & 3318  $\pm$ 13 (46.9\%)\\
\textbf{Failed} & 3381 $\pm$ 12 (47.8\%) & 3497 $\pm$ 11 (49.5\%) & 3754 $\pm$ 13 (53.1\%)\\
\midrule
-  Syntactic & 27 $\pm$ 1 (0.4\%) & 59 $\pm$ 4  (0.9\%)& 103 $\pm$ 3  (1.5\%)\\
-  Semantic & 1303 $\pm$ 23  (18.4\%)& 1415 $\pm$ 15  (20.0\%)& 1182 $\pm$ 18 (16.7\%) \\
-  Operational & 2051 $\pm$ 21 (29.0\%)  & 2023 $\pm$ 6 (28.6\%) & 2468 $\pm$ 16 (34.9\%) \\
\bottomrule
\end{tabularx}

    \end{center}
\end{table}

A high-level results summary is shown in Table~\ref{tab:test-res-llms}. There are a total of 7072 generation tasks in the dataset. All three models are able to solve (generate functions that pass all tests) approximately half of the tasks. Mistral performs worst in our study, solving \tracked{ on average 3318 (46.9\%)} of tasks, and GPT-4o does best solving \tracked{3691 (52.2\%)}, outperforming the worst model by approximately 5 percentage points. 

% Overall, the LLMs had a comparable number of passed tests which make up around half of the total functions. we report the test results of 7072 generated functions in Table \ref{tab:test-res-llms}.

\revised{R1C12}{1}{To get a better idea of whether these results are impacted by the code level of the function, we use the code levels defined by Yu et al. \cite{Yu2024codereval} that are based on the nature of dependencies of the function.} 
The code levels are: self-contained, standard library runnable, public library runnable, class runnable, file runnable, and project runnable. Code levels provide a rough indication of the ``difficulty'' of a generation task, based on what kind of dependencies the LLM needs to correctly incorporate.
% The code levels are: self-contained (no need to import), standard library runnable (no need to install), public library runnable (uses libraries available on PyPI), class runnable (uses code outside the function, but within the class), file runnable (uses code outside the class, but within the file), and project runnable (uses code in other files). Code levels provide a rough indication of the ``difficulty'' of a generation task, based on what kind of dependencies the LLM needs to correctly incorporate.

\begin{figure}[ht!]
    \centering
    \includegraphics[width=0.9\linewidth]{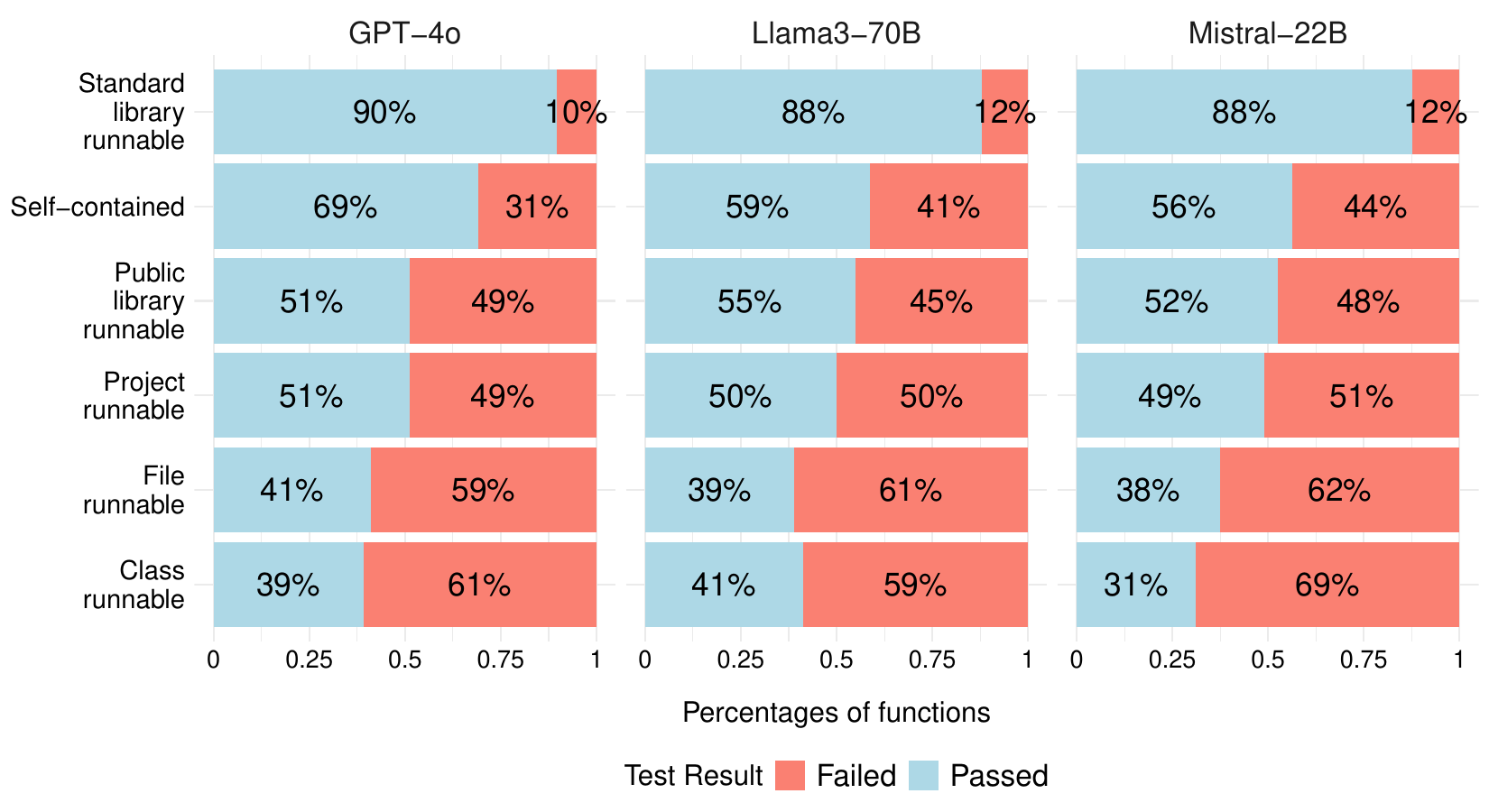}
    \caption{\protect\tracked{Passed and failed functions per LLM for each code level across three runs. The total number of functions per LLM in a single run is 7072.}}
    \label{fig:gpt4-level-test-res}
\end{figure}

We looked into the code levels that passing and failing functions belong to (see Figure \ref{fig:gpt4-level-test-res}). Unsurprisingly, the fail rate for all models increases as tasks get more difficult (i.e., by construction, class runnable tasks tend to be substantially more challenging than self-contained ones, and all models struggle much more with solving them correctly). Pass rates for the easiest type of task (standard library runnable) are close to 90\% for all models, going down to as low as 31\% to 41\% for the most challenging tasks (class runnable). We observe that, overall, all three models perform comparably on most code levels, with the exception of self-contained tests (where GPT-4o outperforms the other models by a larger margin of 10 to \tracked{13} percentage points). This difference explains most of the slightly higher overall performance of GPT-4o. \tracked{We also confirmed these differences using Chi-square test, which assesses the association between categorical variables (code level and pass/fail outcome) resulting in p-value $<$ 0.0001, and Cramér's $\phi$ as an effect size for the relationship between the two nominal variables ($\phi = 0.34$ :- medium effect)}.

% The function distribution among different code levels varied and so the passing and failing rate of the functions that belong to these levels. More specifically, we found that class- and file-runnable functions fail the most, while self-contained and standard lib-runnable functions pass the tests the most. We argue that the reason behind this is that class- and file-runnable functions are more likely to contain local dependencies, and large context (e.g., the whole class) that cannot be covered by the prompting techniques in CodePromptEval and can limit the LLM when generating the functions.

\begin{figure}[!ht]
    \centering
    \includegraphics[width=\linewidth]{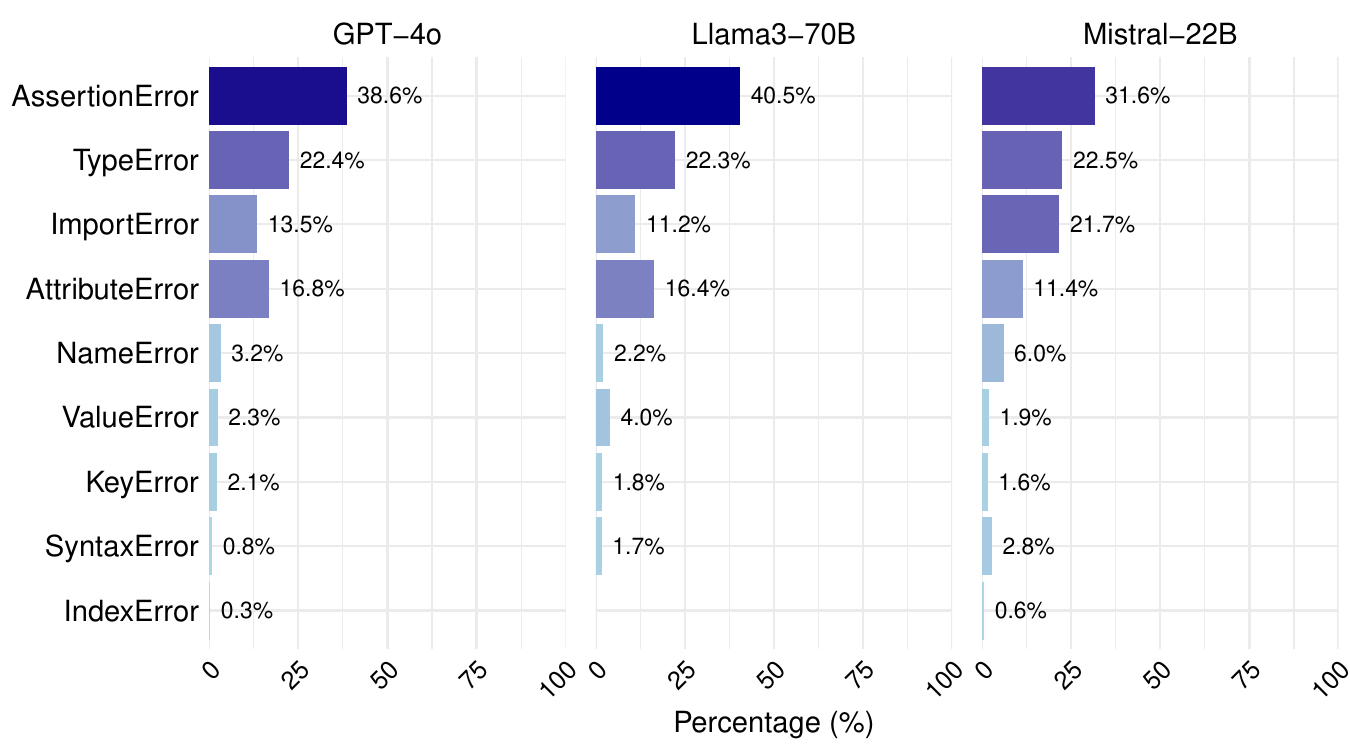}
    \caption{\protect\tracked{Percentages of error types occurring among failing tests for functions generated by GPT-4o, Llama3-70B, and Mistral-22B across three runs.}}
    \label{fig:gpt4-error-types}
\end{figure}

Finally, we report what errors led to the failing tests shown in Table~\ref{tab:test-res-llms} and Figure~ \ref{fig:gpt4-level-test-res}. We report the error types based on the Python exception that is first thrown when running the tests. The results of the error types are visualized in Figure \ref{fig:gpt4-error-types}. The most common error type for failed tests across the LLMs is \texttt{AssertionError}, indicating that the LLM generated a Python function that did not exhibit precisely the expected functionality (as defined through unseen tests). However,  are also frequently encountered such as \texttt{TypeErrors} (operation is performed on a value of an inappropriate type, indicating that the LLM misjudged the runtime type of a Python object), \texttt{AttributeErrors} (invalid attribute reference is made), and \texttt{ImportErrors} (a faulty import of a module or object). Other errors, such as \texttt{NameErrors} or \texttt{IndexErrors}, exist but are rare. While there are differences between the LLMs, they are relatively minor and not systematic. The most notable difference is that Mistral tends to generate functions leading to an \texttt{ImportError} or \texttt{NameError} more frequently than the other LLMs, whereas \texttt{AttributeErrors} are less frequent in Mistral-generated code.

\begin{keyfinding}
\textbf{Key Findings (RQ1):}
Overall, we observed that GPT-4o minorly outperforms the other LLMs in the study. However, in general, results are consistent between current-generation LLMs. Depending on task difficulty, all LLMs can solve between 31\% and 90\% of tasks. Assertion and TypeErrors are the most common cause of failed tests.
\end{keyfinding}

% In addition to reporting the test results of the LLMs' generated functions, we present the specific error types that caused the functions to fail the tests. In general, the most common error type for failed tests across the LLMs is AssertionError (output does not match expected outcome) as illustrated in Figure \ref{fig:gpt4-error-types}. However, TypeErrors (operation is performed on a value of an inappropriate type), AttributeErrors (invalid attribute reference is made), and ImportErrors (a module or object cannot be imported) are also frequently encountered.

% The levels are class-runnable, file-runnable, plib-runnable, self-contained, slib-runnable, project_runnable

\section{Prompt Technique Comparison}
\label{sec:codepromteval-eval}

We now turn towards RQ2, and describe the results of a statistical analysis examining how the different prompt techniques applied in each prompt impact the function regarding (i) correctness, (ii) similarity to the ground truth, and (iii) quality.

\subsection{Correctness}
\label{sec:correctness}

A central question for assessing the value of prompt techniques is how likely a (combination of) techniques is to lead \revised{R1C5}{1}{to a correct code, meaning that it is both (i) syntactically and operationally correct (valid and does not throw errors) and (ii) semantically correct (passes the tests).}
To measure the correctness of different prompts, we use the well-established Pass@k metric \cite{chen2021evaluating}. \revised{R1C7}{1}{This metric measures the likelihood of drawing $k$ passing functions from the results of $n$ number of generations (or repetitions).}
% This metric measures the likelihood that the LLM will generate a correct solution from the $k^{st}$ attempt. When $k=1$, then the metric becomes a measure of accuracy, as it calculates the number of passed functions divided by the total number of functions. 

\begin{figure}[!ht]
    \centering
    \includegraphics[width=0.9\linewidth]{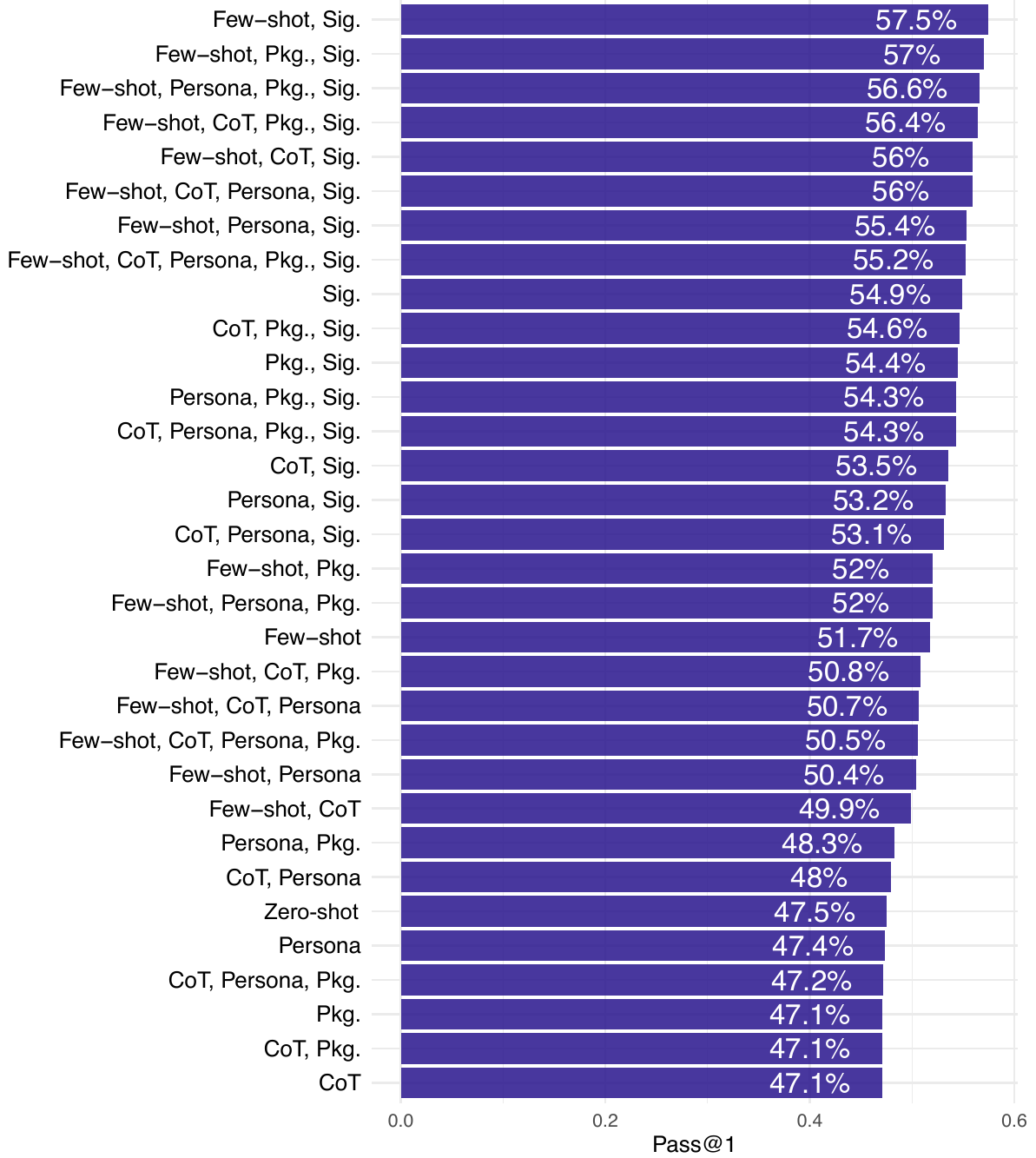}
    \caption{\protect\tracked{Pass@1 results of the different (combinations of) prompt techniques exemplified for GPT-4o.}}
    \label{fig:passk-combs}
\end{figure}

In our study, we run the functions generated by the 32 combinations of prompt techniques \tracked{over three repetitions ($n=3$)} on the tests provided by the CoderEval benchmark. Then, we collect the test results (pass or fail) and measure Pass@1 \tracked{($k=1$)} accordingly.
Figure \ref{fig:passk-combs} shows Pass@1 results for all combinations of techniques (see Table~\ref{tab:combinations}) for GPT-4o. Given that results between different models appear to be very consistent (see also Section~\ref{sec:codepromteval-eval}), we focus our discussion on one example model. However, results for the other models can be found in the supplemental material.
% Appendix \ref{app:passk}.

It is evident from Figure \ref{fig:passk-combs} that the most important technique when it comes to correctness is the presence of a function signature. Combining the signature with other techniques, such as few-shot or chain-of-thought, is sometimes helpful to further increase the likelihood of a generated function being correct (albeit by a very small margin, e.g., adding \tracked{chain-of-thought and few-shot examples to the signature only leads to an improvement of 0.1 percentage points)}. The best combination, with a Pass@1 of 57.5\%, is the combination of signature and few-shot. We achieved the worst results in terms of correctness when using chain-of-thought alone, with a Pass@1 of 47.1\%. It is surprising to note that the impact of prompt engineering techniques is overall lower than we would have expected --- the difference between the best and worst combinations is merely \tracked{10} percentage points, i.e., prompt programming seems to have a noticeable impact in only a little over one in ten generation tasks.

Our findings also indicate that sometimes the addition of more information in the prompt leads to worse performance. For example, using only few-shot and signature performs better than if \tracked{all possible prompt techniques are used}. Further, it is evident that techniques can interact in non-obvious ways. For example, both package information and CoT alone led to the worst Pass@1 results. However, if these techniques are used in conjunction with a function signature, Pass@1 improves marginally over using only the signature in isolation.

% , we show that while there are small differences among the prompt techniques, functions generated via prompts that include Signature are more likely to pass, since Signature was consistently present in all the prompts with the higher $Pass@1$ scores. We also see that the inclusion of Few-shot can also increase the likelihood of the function passing the tests for the different models at different levels.

\begin{figure}[!ht]
    \centering
    \includegraphics[width=0.95\linewidth]{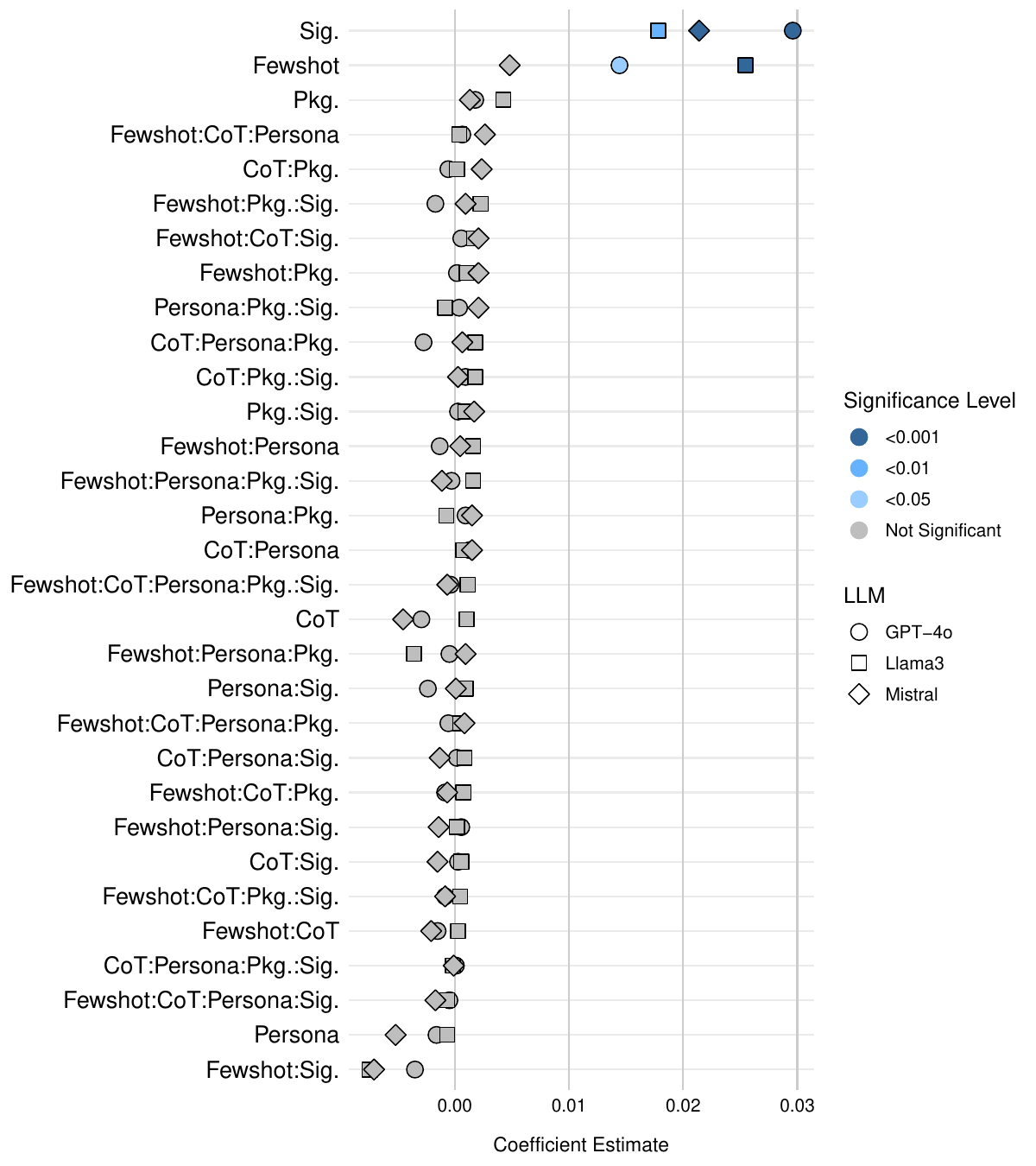}
    \caption{\protect\tracked{Results from our regression analysis for the pass@1 scores. Each point visualizes the coefficient estimate for the corresponding combination. The darker colors represent more conservative significance levels ($\alpha$). Zero-shot is not depicted, as it cannot be combined with other techniques.}}
    \label{fig:test-res-regression}
\end{figure}

To further investigate these interactions between factors in our experiment, we conducted a multi-linear regression analysis. Figure \ref{fig:test-res-regression} shows the five prompt techniques in the study their interactions and their effect on the \tracked{pass@1 score}. For instance, ``CoT:Persona'' describes if the impact on test results comes from the interaction of CoT and Persona in a prompt, regardless of whether that prompt includes other prompt techniques. Similarly, ``Sig.'' (signature) refers to all prompts that include at least the signature (including, for example, P23, the combination of few-shot and signature), and is not limited to prompts that only specify the signature.

The multi-linear regression results in a coefficient estimate and a p-value for each factor and possible interactions among the factors. The coefficient reflects the impact on the test results, positive and negative coefficients refer to positive and negative impacts, respectively. The p-value indicates how significant the impact is.

In line with our previous findings, we observe that the presence of a signature and few-shot in a prompt (regardless of whether they are combined with other prompt techniques) affect the test results positively (albeit with different statistical significance levels for different LLMs), and a positive, high coefficient estimates (meaning there is a significant positive impact on correctness). Interestingly, few-shot does not have a statistically significant impact in the case of the Mistral model.

The remaining main factors (packages, chain-of-thought, and persona) do not have a statistically significant impact on any of the three models. 
% However, it is interesting to observe that the impact of a persona as well as chain-of-thought even trends to the negative (coefficient estimate smaller than 0).

\begin{keyfinding}
\textbf{Key Findings (RQ2.1):}
The presence of a signature or few-shot has the clearest positive impact on correctness. The other prompt techniques in the study do not have a statistically significant impact on correctness. However, in general, the difference between ``good'' and ``bad'' prompt techniques is surprisingly low. Adding additional information to a prompt sometimes leads to worse performance.
\end{keyfinding}

% While the strength of the impact varies across LLMs, we still observe a significant impact of Signature (for all LLMs) and Few-shot (for GPT-4o and Llama3) on the test results. The results also show that while both Few-shot and Signature can improve the test results, the interaction between them (i.e., Few-shot : Signature) does not necessarily affect the test result rather the impact in prompts that contain both Signature and Few-shot techniques comes mainly from the presence of either Few-shot or Signature in the prompt.

% \section{Qualitative Analysis of Errors}
% \label{sec:error-eval}

\begin{figure}[!ht]
    \centering
    \includegraphics[width=0.9\linewidth]{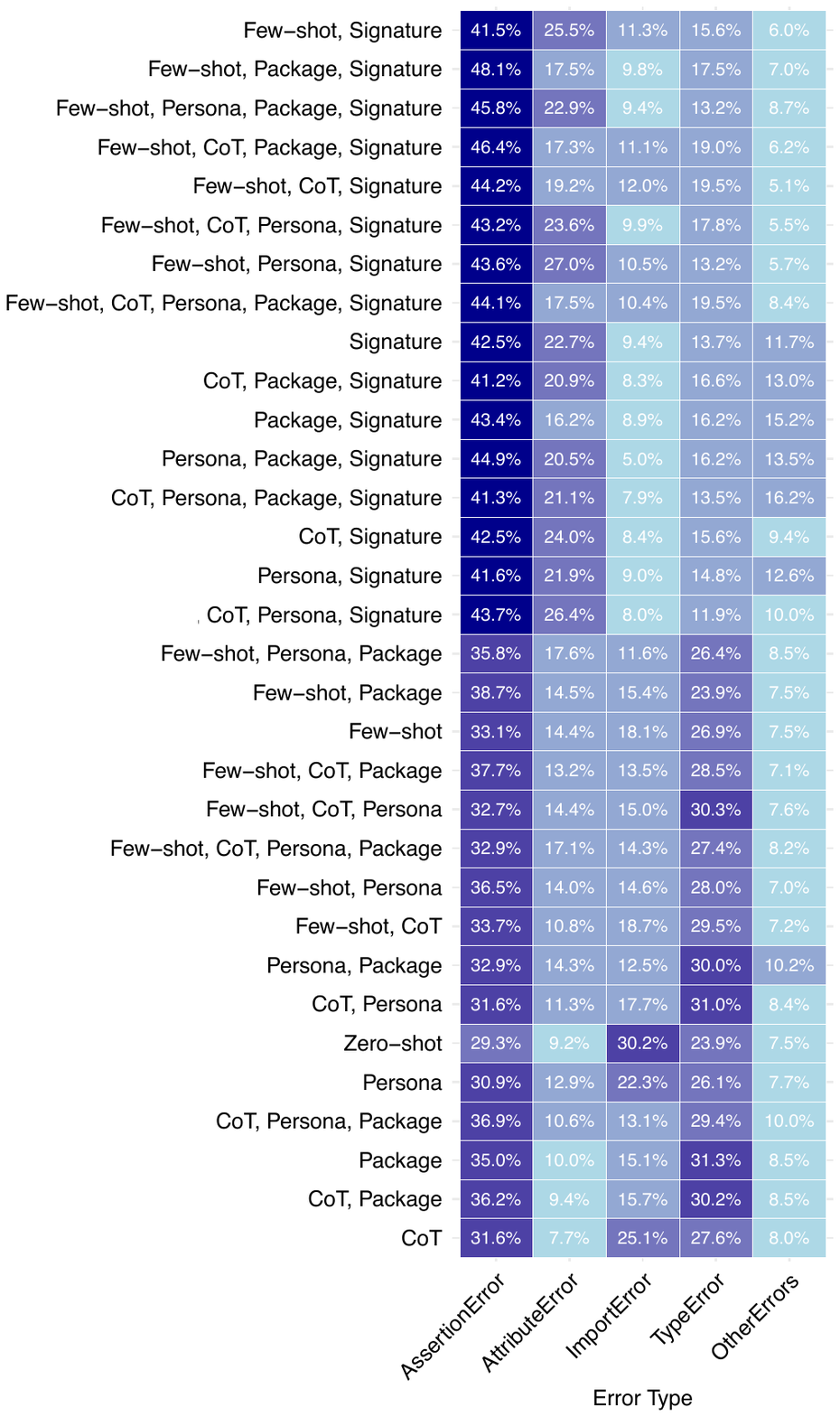}
    \caption{\protect\tracked{The percentages of error types that we observed in failed functions generated by different combinations of prompt techniques (GPT-4o) over three runs.}}
    \label{fig:error-heatmap}
\end{figure}

Digging deeper into what causes generated functions to fail, Figure \ref{fig:error-heatmap} displays the percentages of errors encountered for each combination of prompt techniques. We show the four most common error types (\texttt{AssertionError}, \texttt{TypeError}, \texttt{AttributeError}, and \texttt{ImportError}) using GPT-4o (other models in the supplemental material). For example, \tracked{48.1\%} of the failed functions of prompts with few-shot, packages, and signature throw an \texttt{AssertionError}.
% Appendix \ref{app:error}
% The aim of the heatmap is to show the distribution of the number of failed tests across the error types from a specific combination of prompt techniques rather than to compare the errors among combinations of prompt techniques.

The combinations of prompt techniques are ordered from the fewest errors (at the top) to the most errors (at the bottom). In general, we see that the prompts that result in the least number of errors (among the first rows in the heatmap) are combinations that include a signature. On the other end, the prompts with the most errors lack few-shot examples.

Taking a closer look at the different error types, we observe that while \texttt{AssertionErrors} generally occur at a higher rate than the other error types across all prompt techniques, they are particularly more frequent in prompts that include the function signature\tracked{, meaning that failed functions by prompts with a function signature are able to run but fail their tests due to a semantics-related error}. In contrast, the absence of the function signature often leads to \tracked{\texttt{ImportErrors} as well as \texttt{TypeErrors} that primarily occur because the LLM misjudges the expected number or order of positional arguments when generating functions.}

% \texttt{ImportErrors} were observed more frequently in prompts that did not employ few-shot prompting. 
Interestingly, in a subset of cases, \texttt{ImportErrors} occurred even when packages were explicitly specified. To investigate this, we manually inspected five random prompts where packages were specified but still resulted in \texttt{ImportErrors}. We found that when the prompt indicated the use of a package that is local or unfamiliar to the LLM, the LLM hallucinated and attempted to import non-existent functions from the specified packages.

We note that these findings are consistent for GPT-4o and Llama3, while the errors of code generated by Mistral lacked any clear patterns for the above mentioned error types. 
However, we observed a trend of a higher rate of \texttt{AttributeErrors} in Mistral when the signature is included in the prompt. For the other two LLMs (GPT-4o and Llama3), we did not observe any consistent patterns among the prompts that triggered \texttt{AttributeErrors}.

Overall, we emphasize that encountering a certain error does not necessarily mean that the function is free from the other error types, as the program terminates at the first error thrown. However, assertions are evaluated after the function has successfully been executed, so an \texttt{AssertionError} indeed indicates that no other errors have occurred. Further, \texttt{AssertionErrors} are qualitatively different from other error types, as they do not indicate a fundamentally broken function, but rather that the LLM misunderstood (or could not correctly guess from context) some assumptions about the functionality of the code that is to be generated.

\begin{keyfinding}
\textbf{Key Findings (RQ2.1):}
% Providing the signature or few-shot examples in a prompt can lead to fewer errors in general, and particularly helps to avoid TypeErrors, AttributeErrors, and ImportErrors. Additionally, while providing package information can naturally help reduce some ImportErrors, it can also lead to hallucinated imports if the prompt includes unfamiliar to the LLM.
Including the signature or few-shot examples in prompts generally reduces errors, particularly TypeErrors, AttributeErrors, and ImportErrors. Providing package information can naturally reduce ImportErrors but may cause hallucinated imports if unfamiliar to the LLM.
\end{keyfinding}

\subsection{Similarity}
Beyond correctness, we believe that another important question is how similar generated functions are to the human-written baseline. 
% Note that similarity is inherently neither good nor bad: a high similarity may be good in the sense that if an LLM provides a function that fails the tests but is ``close'' to the baseline it may be simple to repair manually; on the other hand, a low similarity can also be desirable in some cases, e.g., when looking for inspiration or alternative solutions.
\revised{R1C6}{1}{We use the CrystalBLEU score \cite{crystalbleu} to measure how similar the generated function is to the baseline in terms of the syntax and semantics of the function combined. CrystalBLEU is seen as a stricter improvement over the older CodeBLEU metric~\cite{ren2020codebleu}}. 
% It is a composite score that integrates the scores of four n-gram measures (where $n=1,2,3,4$ to measure the syntactic similarity, then uses it as a proxy to estimate the semantic similarity of the generated function to the ground truth.) }
% CodeBLEU is a composite score that integrates the scores of: N-gram match (BLEU \cite{papineni2002bleu}), a weighted BLEU, and the Abstract Syntax Tree (AST) node match, and the dataflow match as a semantic similarity measure with different weights (by default, 25\% for each of the four scores). 
In our analysis, we remove the signature of the generated function and the ground truth before measuring the similarity to avoid any bias toward the signature prompt technique. 
% Consequently, we can no longer measure the dataflow similarity as the functions are no longer parsable (i.e., identifying parameters and their dataflow in the function). Therefore, we set the weight for the dataflow match to 0, and 1/3 for the three other match scores.

\begin{figure}[!ht]
    \centering
    \includegraphics[width=0.9\linewidth]{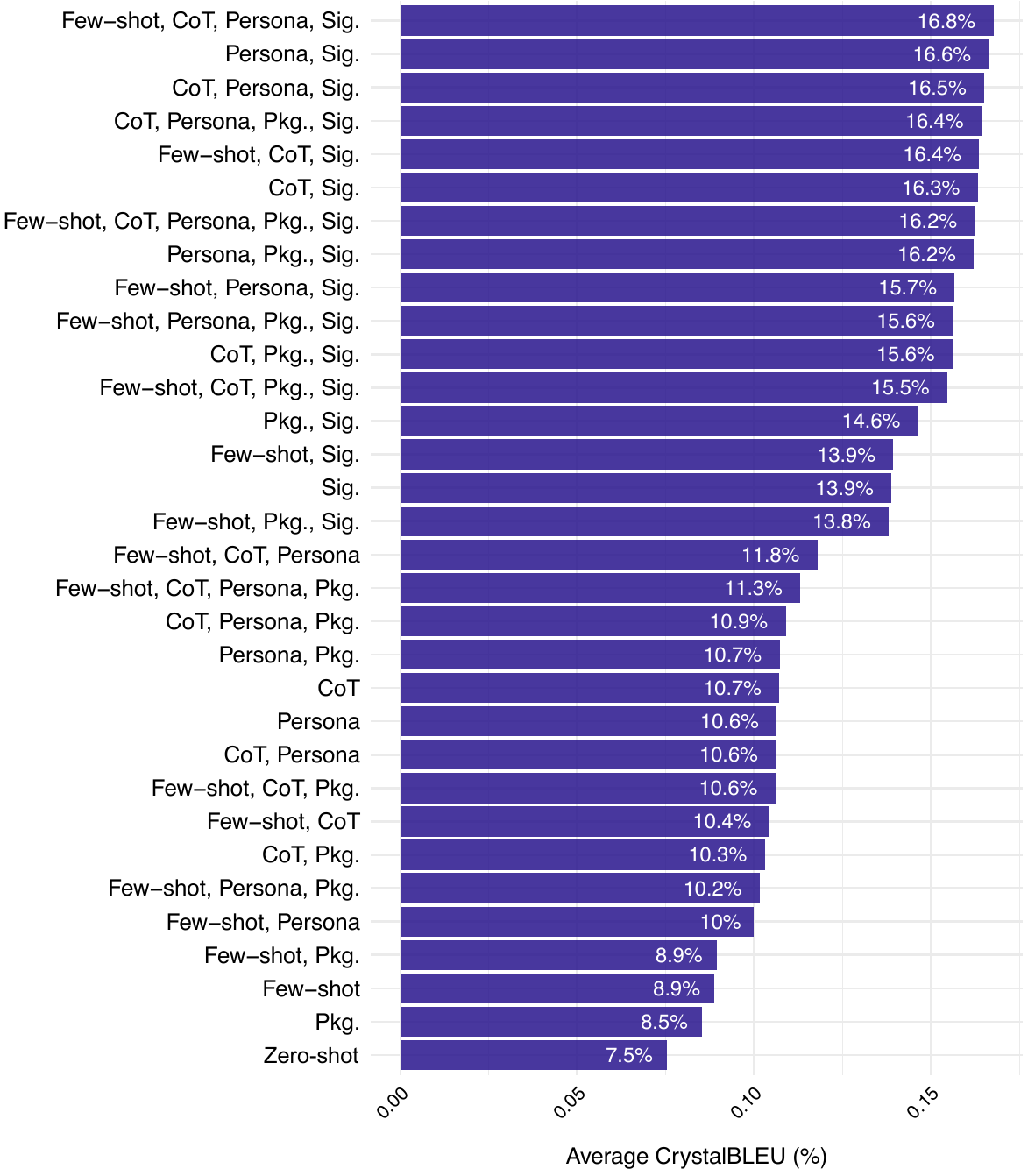}
    \caption{\protect\tracked{The average CrystalBLEU scores for the functions generated by each combination (Llama3).}}
    \label{fig:codebleu-avg-llama3}
\end{figure}

\begin{figure}[!ht]
    \centering
    \includegraphics[width=0.95\linewidth]{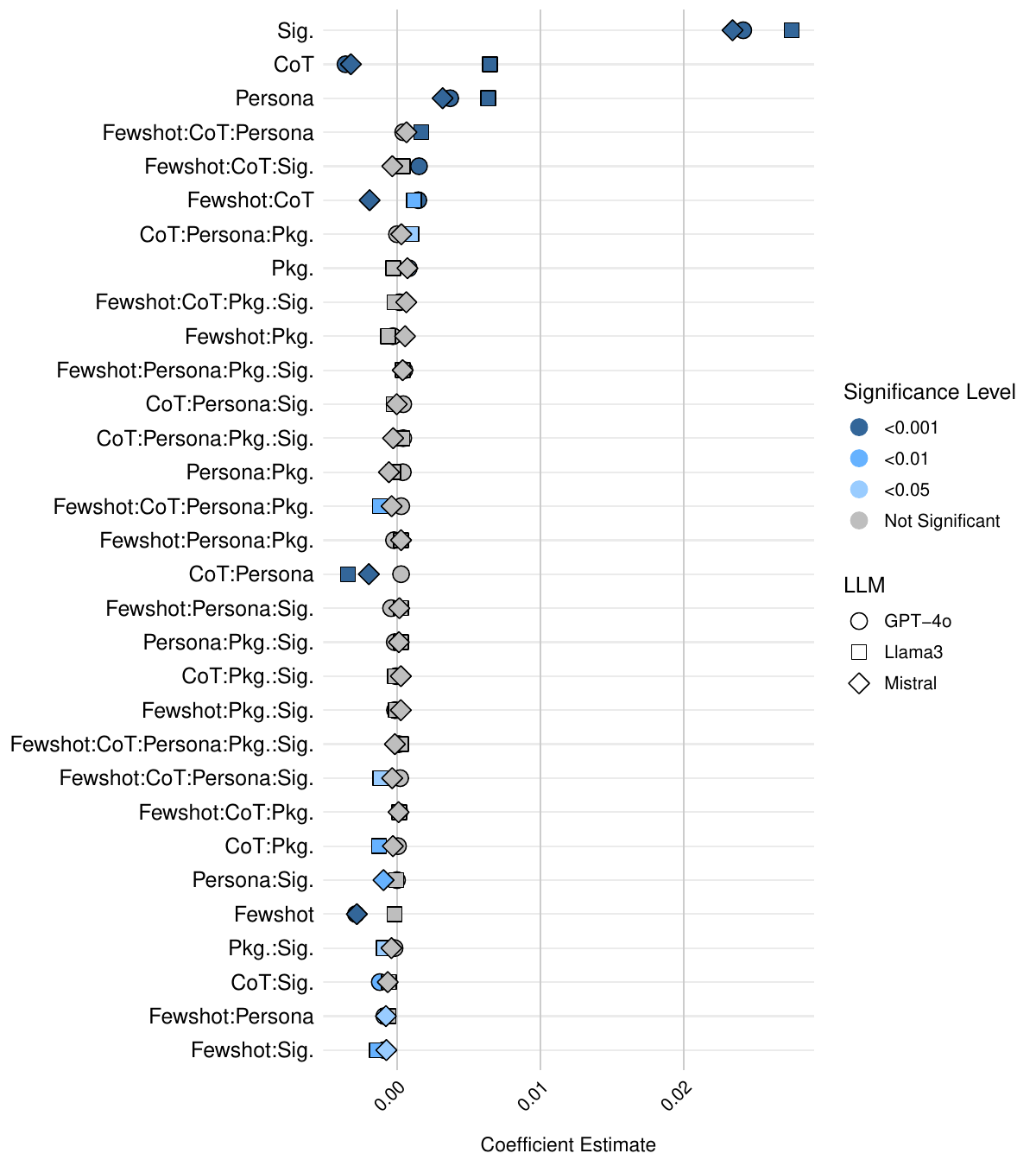}
    \caption{\protect\tracked{The coefficient estimates from the linear mixed-effects regression of prompt technique combinations that significantly impact the CrystalBLEU score.}}
    % Appendix~\ref{app:sim}
    \label{fig:sim-scores-regression}
\end{figure}

In Figure \ref{fig:codebleu-avg-llama3}, \tracked{we see that the average CrystalBLEU score across three runs is low for all approaches (varying between 16.8\% and 7.5\% for Llama3)} indicating that generated solutions are largely different than how humans have solved the same tasks. Results for the other models are in the supplemental material.

We observe that using any prompt technique increases similarity (i.e., zero-shot has the lowest similarity to the baseline for all three models).
Consistently with correctness, combinations that include a signature lead to higher similarity. This is unsurprising, given that a predefined signature restricts the solution space for the LLM (which can be desirable or unwanted depending on context). Combining more techniques indeed seems to generally increase similarity.
\tracked{We also observe that few-shot can decrease the similarity, achieving a score of 8.5\%. However, combining it with signature, chain-of-thought or persona can improve the similarity to 10\% and above}. 

To better understand the impact of the prompt techniques and their interactions on the code similarity, \tracked{we now perform a linear mixed-effects regression analysis to see how the different prompt techniques and their interactions can impact the CrystalBLEU while accounting for the within-group variation (random effects) that arise from the three runs of each LLM.}

% N-gram and weighted N-gram matches can give us insights into how similar the generated function to the baseline on a lexical level, i.e., the set of words such as coding style, variable naming, and/or in-line documentation in the function. The AST match focuses on the syntactic structure and code constructs in the function such as nested statements independently of their naming (e.g., loops, conditionals, operations).

Figure \ref{fig:sim-scores-regression} \tracked{shows our linear mixed-effects regression results. We see that, regardless of the test results, the presence of a signature or persona in a prompt can significantly increase the CrystalBLEU score. Chain-of-thought (CoT) seems to also positively impact the CrystalBLEU score for Llama3, but significantly lower it for GPT-4o and Mistral.}
% \tracked{We also confirm our observations in Figure \ref{fig:codebleu-avg-llama3} regarding the impact of few-shot. For selected models, the inclusion of few-shot examples in a prompt can lower the similarity until it is combined with a prompt technique with a positive impact such as CoT or persona, which results in a positive interacting impact of few-shot together with CoT alone or with a persona for selected models.}
\tracked{We also observed that the interaction between certain prompt techniques can either reinforce or counteract the effects seen when the techniques are used individually. For example, while both Signature and CoT independently increase similarity in Llama-generated functions, using them together in a prompt can reverse that positive effect and significantly reduce similarity.}

% In an interesting case, ... maybe I can talk about the Cot:Sig thing for Llama3??

% The impact of the signature was shared across all dimensions of CodeBLEU. The impact of the persona and CoT was only observed for the lexical similarity, which explains the slightly lower $p$-values compared to the signature.
% In an interesting case, few-shot seems to lower the lexical similarity but increase the syntactic similarity, which results in no overall impact on the CodeBLEU. The impact of few-shot suggests that while it generates functions that may use different variable naming and style (lower lexical similarity), the structure and logic are close to the baseline (higher syntactic similarity).

\begin{keyfinding}
\textbf{Key Findings (RQ2.2):}
\tracked{The signature and persona increase the overall similarity of the function to the baseline (i.e., code written by humans). Few-shot decreases the similarity unless combined with chain-of-thought.}
% shorter version
% The signature, persona, and CoT increase the overall similarity of the function to the ground truth code. Few-shot increases the syntactic similarity (function structure) and decreases the lexical similarity (variable naming).
\end{keyfinding}

\subsection{Quality}

Using prompt techniques that yield correct functions does not necessarily mean that these functions are maintainable and of good quality. Hence, we now turn to an assessment of the quality of the generated code. In our experiment, we focus on code smells and complexity as proxies of code quality. For this analysis, we only evaluate functions that \emph{pass} their tests (see Section~\ref{sec:correctness}). We do not believe that assessing the quality of functionally incorrect implementations is fruitful because refactoring must be done on a working piece of code and preserve its behavior~\cite{fowler2018refactoring}.
% to evaluate the maintainability of the \textit{passing} functions generated by different combinations of prompt techniques in CodePromptEval.

For code smells, we run Pylint on the generated functions for each prompt in CodePromptEval, using the code smell IDs defined by Pylint.
%\footnote{\url{https://pylint.readthedocs.io/en/stable/user_guide/messages/messages_overview.html#}}. 
Then, we group the code smells for prompts that share the same prompt techniques, and finally, we \tracked{select the top 15 code smells that were the most frequent across all prompt techniques.}
% filter code smells that make up less than 5\% of the total number of code smells for the group.

\begin{table}[!ht]
\scriptsize
\centering
\caption{List of Code Smells in generated functions.}
\label{tab:code-smells-mapping}
\begin{tabularx}{\linewidth}{lcl}
\toprule
\textbf{Category} & \textbf{Code Smell ID} & \textbf{Definition} \\
\midrule
Error & E0602 & Usage of an undefined variable. \\
% Warning & W0212 & Access to a protected class member. \\
Warning & W0611 & Import statement not used. \\
Warning & W0613 & Function argument is not used. \\
% Warning & W0621 & Redefines name from outer scope. \\
% Warning & W1203 & Improper string formatting in logging. \\ % Llama3 does not have this, so I removed it
Refactoring & R0903 & Insufficient public methods in a class. \\
Refactoring & R1705 & Unnecessary ``else'' after ``return''. \\
% \tracked{Refactoring} & \tracked{R0913} & \tracked{Too many arguments.} \\
Convention & C0301 & Line exceeds the character limit. \\
Convention & C0103 & Violating UPPER\_CASE naming style. \\
Convention & C0115 & Class lacks a descriptive docstring. \\
Convention & C0116 & Function lacks a descriptive docstring. \\
\tracked{Convention} & \tracked{C0411} & \tracked{Wrong import order.} \\
Convention & C0304 & File missing a final newline. \\
\bottomrule
\end{tabularx}
\end{table}

We find \tracked{11} code smells that fulfill these criteria for all LLMs (see Table \ref{tab:code-smells-mapping}). Most identified code smells are \textit{convention} code smells, but there is also one \textit{error}\tracked{, two warnings,} and two \textit{refactoring} smells.
From this list, we decided to remove C0304 as it is present in all generated functions across all LLMs and is mostly an artifact of our generation pipeline.

% We found that all the LLMs shared 12 main code smells (see ), the code smells mainly belong to the \textit{warnings} and Python \textit{conventions} categories, but there were also one \textit{error} and a couple of needed \textit{refactoring} actions.

\begin{figure}[!ht]
    \centering
    \includegraphics[width=\linewidth]{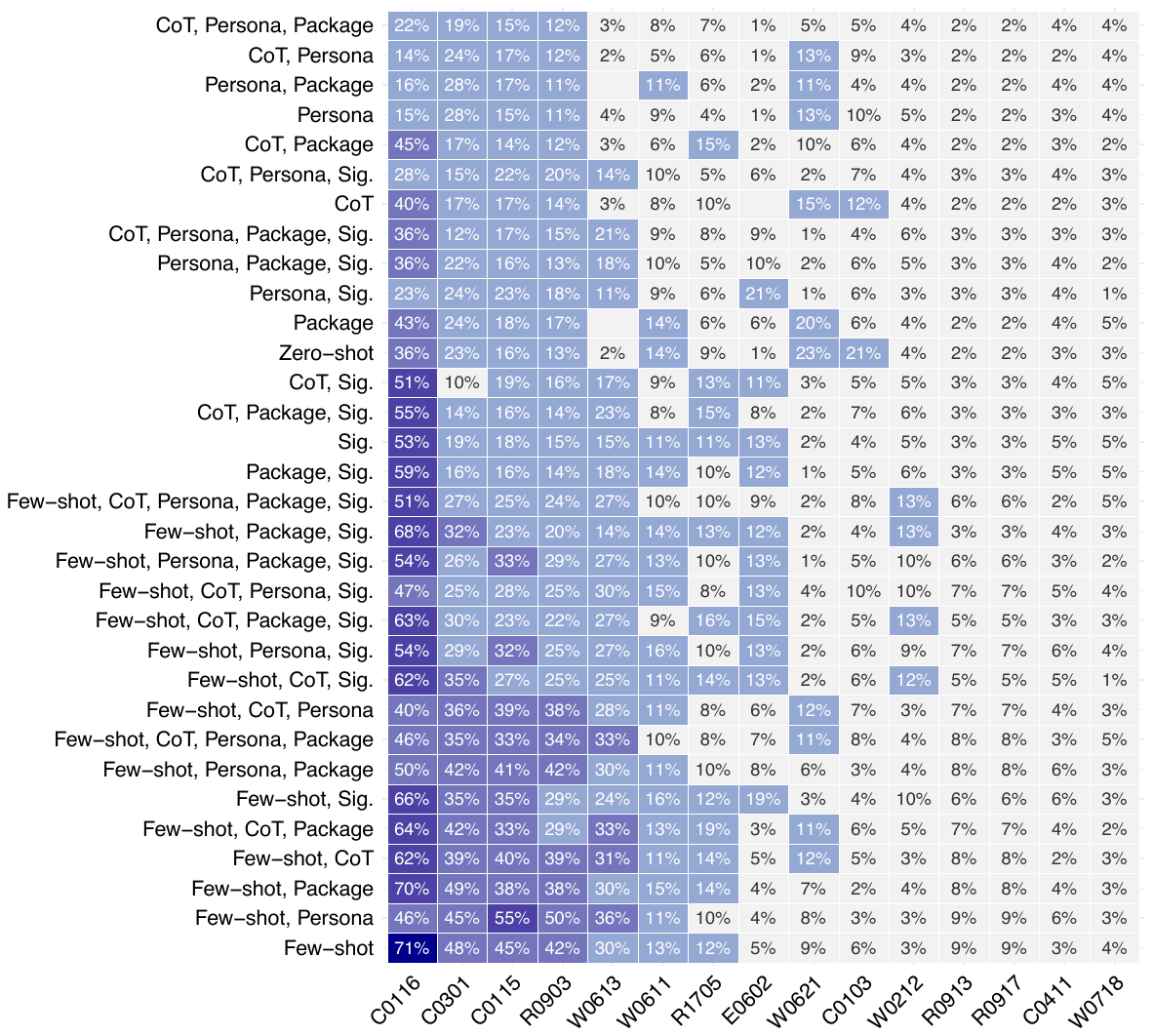}
    \caption{\protect\tracked{Percentages of functions generated by GPT-4o that have different code smells. Empty fields indicate that no smell of this type is found. Note that functions can have instances of multiple types of smells.}}
    \label{fig:lints-gpt4}
\end{figure}

In Figure \ref{fig:lints-gpt4}, we show what percentage of functions have at least one instance of each code smell. For reasons of brevity, we focus on GPT-4o (Llama3 and Mistral's in the supplemental material).
% Appendix \ref{app:lints}

We note that \tracked{71\%} of the functions generated using the few-shot technique contain C0116 code smells, indicating that these functions lacked a descriptive docstring (in contrast to only \tracked{22\%} of functions generated by chain-of-thought combined with a persona and package information).
In general, we observe that prompts that apply the few-shot and signature prompt techniques generate functions with more code smells and, more specifically, warning and error code smells compared to other prompts.

On the other hand, we observed that CoT, persona, and package lead to functions with fewer code smells, unless these prompt techniques are combined with few-shot and/or signature, then the percentage of code smells increases. This is interesting, as we have seen that few-shot and signature are the techniques with the clearest positive impact on correctness (see Section~\ref{sec:correctness}). In part, this discrepancy could be explained by solutions for challenging tasks that LLMs only solve correctly when provided examples or a signature (recall that, for this analysis, we have only investigated functions that pass all tests --- hence, some challenging functions have an analyzable solution for signature and few-shot, but not other techniques). However, we note that the differences in Figure \ref{fig:lints-gpt4} are too large to be entirely explained in this way. Consequently, we conclude that CoT, persona, and package information indeed seem to systematically lead to fewer code smells.

\begin{keyfinding}
\textbf{Key Finding (RQ2.3):}
While using CoT, persona, or package information leads to fewer correct solutions, these techniques lead to higher-quality code in terms of code smells.
\end{keyfinding}

% we observed that compared to 265 total code smells of functions generated with Zero-shot (No prompt technique implemented), Few-shot and Signature increase the number of code smells up to 451 code smells. On the other hand, the use of Persona and/or Chain-of-Thought prompt techniques in the prompt can minimize the number of code smells. Specifying the Packages in the prompt also seems to reduce the number of code smells slightly. 

We now turn towards the cyclomatic and cognitive complexity and compare the complexity of generated solutions to the complexity of the human-written baseline.
% (using both, cyclomatic and cognitive complexity).
% Our overall approach for this analysis is to compare the complexity of generated solutions to the complexity of the human-written baseline.
In Table \ref{tab:cyclo-all}, we show the $p$-values resulting from the \textit{paired} Wilcoxon test to assess the statistical significance of differences between the cyclomatic complexity of the generated functions and ground truth. We only show the prompt techniques that had a significant impact on the complexity for at least two LLMs ($\alpha = 0.05$). Complete results are in the supplemental material.

% (we consider a level of significance $\alpha = 0.05$).

\begin{table*}[!ht]
\centering
\scriptsize
\caption{\protect\tracked{Cyclomatic complexity analysis using Wilcoxon test and A12 Vargha Delaney for the effect size (N- Negligible, S- Small, M- Medium, L- Large). $\downarrow$ indicates a reduction in complexity, $\emptyset$ indicates no statistical difference.}}
\begin{tabularx}{0.68\textwidth}{l|rrlrrlrrl}
\toprule
\textbf{Combinations} & \multicolumn{3}{c}{\textbf{GPT-4o}} & \multicolumn{3}{c}{\textbf{Llama3}} & \multicolumn{3}{c}{\textbf{Mistral}} \\
\cmidrule(lr){1-1} \cmidrule(lr){2-4} \cmidrule(lr){5-7} \cmidrule(lr){8-10}
 & \textbf{p-value} & \textbf{A12} & \textbf{Effect} &  \textbf{p-value} & \textbf{A12} & \textbf{Effect} &  \textbf{p-value} & \textbf{A12} & \textbf{Effect} \\
 
\cmidrule(lr){2-4} \cmidrule(lr){5-7} \cmidrule(lr){8-10}
% shared - 3 LLMs
Package            & \textbf{0.0015} & \textbf{0.403} & \textbf{$\downarrow$ (S)} & \textbf{0.0017} & \textbf{0.410} & \textbf{$\downarrow$ (S)} & \textbf{0.0050} & \textbf{0.412} & \textbf{$\downarrow$ (S)} \\
Persona, Package       &\textbf{0.0136} & \textbf{0.442}& \textbf{$\downarrow$ (S)} & \textbf{0.0033} & \textbf{0.412} & \textbf{$\downarrow$ (S)} & \textbf{0.0046} & \textbf{0.424} & \textbf{$\downarrow$ (S)} \\
Zero-shot          & \textbf{0.0172} & \textbf{0.445}& \textbf{$\downarrow$ (S)}  & \textbf{0.0009} & \textbf{0.401} & \textbf{$\downarrow$ (S)} & \textbf{0.0001} & \textbf{0.384} & \textbf{$\downarrow$ (S)} \\
CoT, Package          & \textbf{0.0188} & \textbf{0.448}& \textbf{$\downarrow$ (S)}  & \textbf{0.0118} & \textbf{0.442} & \textbf{$\downarrow$ (S)} & \textbf{0.0016} & \textbf{0.409} & \textbf{$\downarrow$ (S)} \\
CoT, Persona, Package   & 0.0213 & 0.451 & $\downarrow$ (N)  & \textbf{0.0015} & \textbf{0.431} & \textbf{$\downarrow$ (S)} & \textbf{0.0067} & \textbf{0.433} & \textbf{$\downarrow$ (S)} \\

% 2 LLMs - Llama3 and Mistral 
Persona, Sig.      & 0.0755 & 0.475 & $\emptyset$  & \textbf{0.0230} & \textbf{0.440} & \textbf{$\downarrow$ (S)}   &  \textbf{0.0041} & \textbf{0.442} & \textbf{$\downarrow$ (S)} \\
Persona            & 0.0524 & 0.466 & $\emptyset$  & \textbf{0.0015} & \textbf{0.401} & \textbf{$\downarrow$ (S)}   &  \textbf{0.0004} & \textbf{0.408} & \textbf{$\downarrow$ (S)} \\
Sig.             & 0.0909 & 0.460 & $\emptyset$ & \textbf{0.0093} & \textbf{0.430} & \textbf{$\downarrow$ (S)}   &  \textbf{0.0076} & \textbf{0.444} & \textbf{$\downarrow$ (S)} \\

% all 3 but 2 are Neg.
Package, Signature   & 0.0295 & 0.460 & $\downarrow$ (N)  & \textbf{0.0398} & \textbf{0.436} & \textbf{$\downarrow$ (S)} & 0.0143 & 0.458 & $\downarrow$ (N) \\

\bottomrule
\end{tabularx}
\label{tab:cyclo-all}
\end{table*}

We use Vargha Delaney A12 measure \cite{vargha2000critique} to understand the nature of the impact (reduces or increases complexity) and to quantify the effect size (Negligible ($A_{12} \geq 0.45$), Small ($0.36 \leq A_{12} < 0.45$), Medium ($0.29 \leq A_{12} < 0.36$), or Large ($A_{12} \leq 0.29$)). Vargha Delaney A12 is a probability measure (that was later adopted as an effect size measure), which describes the probability that one level (generated function complexity) is greater than a corresponding value in another level (ground truth complexity). If the A12 is less than 0.50, it means that the values of the first level are lower than the second level, and the lower the score is, the larger the effect size.
This allows us to see if the prompts generate functions with a significantly lower or higher complexity as the ground truth, or with a comparable complexity when no significance is observed.
% We also quantify the effect size using Vargha and Delaney's A12 statistic, and represent the magnitude effect sizes for the significant complexity reduction as Negligible ($A_{12} \geq 0.45$), Small ($0.36 \leq A_{12} < 0.45$), Medium ($0.36 \leq A_{12} < 0.29$), or Large ($A_{12} \leq 0.29$) \cite{vargha2000critique}.
% Appendix \ref{app:comp}.
% We highlight the combinations that at least had a small effect size across two or more LLMs.

\begin{table*}[!ht]
\centering
\scriptsize
\caption{\protect\tracked{Cognitive complexity analysis using Wilcoxon test and A12 Vargha Delaney for the effect size (N- Negligible, S- Small, M- Medium, L- Large). $\downarrow$ indicates a reduction in complexity, $\emptyset$ indicates no statistical difference.}}
\begin{tabularx}{0.7\textwidth}{l|rrlrrlrrl}
\toprule
\textbf{Combinations} & \multicolumn{3}{c}{\textbf{GPT-4o}} & \multicolumn{3}{c}{\textbf{Llama3}} & \multicolumn{3}{c}{\textbf{Mistral}} \\
\cmidrule(lr){1-1} \cmidrule(lr){2-4} \cmidrule(lr){5-7} \cmidrule(lr){8-10}
 & \textbf{p-value} & \textbf{A12} & \textbf{Effect} &  \textbf{p-value} & \textbf{A12} & \textbf{Effect} &  \textbf{p-value} & \textbf{A12} & \textbf{Effect} \\
\cmidrule(lr){2-4} \cmidrule(lr){5-7} \cmidrule(lr){8-10}
% GPT-4 and Llama3
    Few-shot, Persona & \textbf{0.0239} & \textbf{0.437} & \textbf{$\downarrow$ (S)}    & \textbf{0.0106} & \textbf{0.403} & \textbf{$\downarrow$ (S)}  & 0.2629 & 0.461 & $\emptyset$ \\
    Few-shot, Persona, Package & \textbf{0.0326} & \textbf{0.445} & \textbf{$\downarrow$ (S)}    & \textbf{0.0210} & \textbf{0.419} & \textbf{$\downarrow$ (S)}  & 0.8239 & 0.505 & $\emptyset$ \\

% Llama3 and Mistral
    Persona, Sig. & 0.1846 & 0.473 & $\emptyset$ & \textbf{0.0024} & \textbf{0.417} & \textbf{$\downarrow$ (S)} & 0.0438 & 0.465 & $\downarrow$ (N) \\
    CoT, Persona & 0.4080 & 0.524 & $\emptyset$ & \textbf{0.0212} & \textbf{0.441} & \textbf{$\downarrow$ (S)} & \textbf{0.0052} & \textbf{0.445} & \textbf{$\downarrow$ (S)} \\
    CoT, Package & 0.1694 & 0.485 & $\emptyset$ & \textbf{0.0245} & \textbf{0.448} & \textbf{$\downarrow$ (S)} & \textbf{0.0334} & \textbf{0.435} & \textbf{$\downarrow$ (S)} \\
    CoT & 0.3593 & 0.516 & $\emptyset$  & \textbf{0.0444} & \textbf{0.458} & \textbf{$\downarrow$ (S)} & \textbf{0.0093} & \textbf{0.432} & \textbf{$\downarrow$ (S)} \\
    Persona & 0.2361 & 0.513 & $\emptyset$   & \textbf{0.0027} & \textbf{0.398} & \textbf{$\downarrow$ (S)}  & \textbf{0.0155} & \textbf{0.442} & \textbf{$\downarrow$ (S)} \\
    Package, Sig. & 0.1617 & 0.468 & $\emptyset$  & \textbf{0.0003} & \textbf{0.406} & \textbf{$\downarrow$ (S)} & \textbf{0.0437} & \textbf{0.448} & \textbf{$\downarrow$ (S)} \\
    Package & 0.1694 & 0.485 & $\emptyset$  & \textbf{0.0006} & \textbf{0.383} & \textbf{$\downarrow$ (S)}  & \textbf{0.0267} & \textbf{0.426} & \textbf{$\downarrow$ (S)} \\
    Zero-shot & 0.2522 & 0.504 & $\emptyset$  & \textbf{0.0008} & \textbf{0.385} & \textbf{$\downarrow$ (S)}  & \textbf{0.0298} & \textbf{0.415} & \textbf{$\downarrow$ (S)} \\

\bottomrule
\end{tabularx}
\label{tab:cogn-all}
\end{table*}

Similar to the code smells results, we see that CoT, persona, and packages reduce the complexity in comparison to the baseline. A zero-shot prompt also leads to lowered complexity. However, all reductions have (at most) a small effect size. This can be explained by the low cyclomatic complexity of all LLM tasks --- in general, only minor simplifications are even possible to the generally rather simple code snippets.

For cognitive complexity (see Table~\ref{tab:cogn-all}), we observe larger differences among the LLMs than between combinations of prompt techniques in general. There was \tracked{no} combination of prompt techniques that reduced the cognitive complexity across all three LLMs. GPT-4o seems to generate functions with no or \tracked{small} differences to the ground truth. Mistral can reduce the cognitive complexity with a small effect size when the prompt does not include few-shot and a persona, packages or CoT applied in the prompt.
In contrast, there are no clear trends or patterns among the prompts in Llama3 rather most of the prompt techniques seem to reduce the cognitive complexity with a small effect size. We conclude that Llama3 appears to lead to simpler solutions than the other models, particularly GPT-4o.

It is interesting to observe that no combination of prompt techniques leads to \emph{more complex} solutions than the baseline --- generated solutions are always slightly simpler or comparably complex. Viewed positively, this may indicate that LLMs generate rather clean code. However, a more negative interpretation may also be that the generated code does not cover some complex corner cases that human-written solutions account for (which may not be covered by CoderEval tests).

\begin{keyfinding}
\textbf{Key Findings (RQ2.3):}
There are noticeable differences among models with regard to the complexity of the code they produce. 
Llama3 appears to produce simpler solutions systematically. 
There were no cases of increased complexity --- LLM solutions were comparably complex to human-written code, or simpler.
\end{keyfinding}

\section{Discussion}
In this section, we discuss the key lessons learned from this study, the implications of our findings for software engineering practitioners and researchers, as well as validity threats.
% We now discuss the lessons learned, the implications of our findings for practitioners and researchers, and validity threats.

\subsection{Lessons learned}

\textbf{L1: The differences in the results of prompt techniques are not dramatic: }
We carefully designed a full factorial experiment to evaluate not only prompt techniques but also combinations of them in a prompt. Our analyses revealed that, while there was an impact of some prompt techniques on the generated functions, the results for most of the prompt techniques were not that different. For example, the difference in the Pass@1 rates for the prompts with the highest and lowest rates is only around 10 percentage points (see Figure \ref{fig:passk-combs}), and the effect sizes of the complexities are mostly small or insignificant (see Table \ref{tab:cyclo-all}). These insights align with other studies that evaluate prompt techniques on code summarization \cite{wang2024advanced} and generation \cite{wang2024enhancing}, where the performance results of different prompt techniques such as CoT, few-shot, self-collaboration, among others, are also comparable.
In contrast, we see clearer differences in the performance results of some prompt techniques when using benchmarks for math-related tasks or general question-answering \cite{cot, kojima2022large}. We conclude that a strong emphasis on prompt programming is not necessary in the context of function-level code generation using current-generation models.

\textbf{L2: Providing information about the interface via few-shot or signature is useful, but limits the ``creativity'' of the LLM: }
In our correctness and similarity results, the signature and few-shot prompt techniques stood out among other prompt techniques. In general, we believe that while they are two different prompt techniques, they can provide similar context about the expected functional interface in terms of positional arguments and expected output. This was also revealed through our general multi-linear regression results in Figures \ref{fig:test-res-regression} and \ref{fig:sim-scores-regression}, where we see that having either signature or few-shot examples significantly impact the code's correctness or similarity, but their interaction or combination does not help. In relation to previous work by Ahmed et al. \cite{ahmed2022multilingual, ahmed2024automatic}, we observe a similar pattern where contextual information about the parameters and other identifiers can improve the code summarization. However, providing this information limits the solution space for the LLM (i.e., it restricts the potential for ``creativity''), which may not always be desired.

\textbf{L3: There is a trade-off between correctness and maintainability when choosing prompt techniques: }
% We have seen opposing results in our analysis of correctness and code quality. On one hand, the prompts that generated the most correct functions included few-shot examples or the function signature, while the prompts that scored the lowest passing rates often applied CoT, persona, and/or package information. On the other hand, we found that prompts that employ mainly the persona, but also CoT and packages significantly improve the maintainability of the generated code (see Tables \ref{tab:cyclo-all},\ref{tab:cogn-all} for complexity and Figure \ref{fig:lints-gpt4} for code smells). 
Our analysis revealed contrasting results: prompts with few-shot examples or function signatures improved correctness but increased complexity and number of code smells, while prompts that employed persona, CoT or package had lower passing rates but significantly enhanced code maintainability (see Tables \ref{tab:cyclo-all},\ref{tab:cogn-all} for complexity and Figure \ref{fig:lints-gpt4} for code smells).
While previous research suggests that the use of a persona in the prompt does not improve the outcome \cite{zheng2024persona} but can improve the personalization and user experience \cite{tseng2024persona}, we believe that this only applies to simple personas such as ``software developer''. However, our results indicate that personas can be more beneficial when used as a way to induce additional quality requirements e.g., ``software developer who writes clean and simple code''. Recent work has also shown that personas can be beneficial for code generation when used in more complex approaches such as self-collaboration where multiple personas (e.g., requirement engineer, software tester, and a developer) are used together to iteratively construct the code in a systematic way \cite{dong2024collaboration}.

\subsection{Implications}
% implications
\textbf{I1: Researchers should prioritize refining prompts for more effective prompt programming experiments}
We shed light on two components of experiments in prompt programming: the generation tasks, and the prompts. Based on observations in our previous work \cite{khojah2024beyond}, we note that developers often use LLMs for more complex tasks than those in common datasets such as HumanEval \cite{athiwaratkun2022multihumaneval} or CoderEval \cite{Yu2024codereval}. Although we were able to analyze and compare different prompt techniques, we believe that a dataset with are more representative functions of the large systems and projects that developers typically work with is needed.

Further, prompts in common benchmarks often lack a consistent format or level of detail. For instance, the prompts in CoderEval are based on functions' docstrings rather than actual prompts. Sclar et al. \cite{sclar2023quantifying} show that LLMs, regardless of their sizes and number of parameters, are highly sensitive to small prompt changes such as prompt formatting. We observed similar behavior when experimenting with the template \emph{``The function uses the following packages''} for the packages prompt technique and found that it caused errors related to using the wrong packages. We traced the issue back to the prompt itself and realized that the packages listed were not necessarily used by the function but existed in its class. When we modified the template to \emph{``The function has access to (but does not necessarily use) the following packages,''} we mitigated the issue. This pre-processing of the prompt technique templates is another aspect of prompt programming recommended by Obrien et al. \cite{obrien2024todo}.
%Note that the issue was not with the prompt technique but rather the formulation of the prompt. 
We found value in inspecting and refining prompts and creating our own few-shot examples, which increased our confidence in the dataset's reliability and stability. Therefore, we encourage researchers to invest in similar efforts.

\textbf{I2: Software developers should avoid overusing prompt techniques}
While we saw that prompt techniques can be beneficial for certain criteria (e.g., signature for correctness and persona for quality), we also saw that combining them does not necessarily yield better results. In fact, some cases showed that including an additional prompt technique can cancel out the impact of the existing prompt techniques. For instance, in the code smells results in Figure \ref{fig:lints-gpt4}, we show how the inclusion of few-shot examples to CoT and persona can increase the code smells by more than one-third. Previous work has also shown how few-shot examples can hurt the LLM performance if not carefully engineered by humans \cite{kojima2022large}.
% In general, we recommend using simple prompts, to avoid confusing the LLM with unnecessary context and requirements. 

\textbf{I3: Different LLMs have different sensitivity levels to the prompt techniques}
We argue that a single prompt technique does not have the same impact on the different aspects (correctness, similarity, or quality) of the generated code across all LLMs. We have seen that the three LLMs demonstrated different sensitivity levels to the prompt.
% We argue that selecting a suitable LLM by understanding how sensitive LLMs are to prompt techniques can be more crucial than investing effort in finding the ``perfect prompt''. 
For instance, we saw how the similarity scores of Llama3-generated functions were significantly impacted by CoT, while it showed no effect for GPT-4o and Mistral (see Figure \ref{fig:sim-scores-regression}).
The error types for Mistral 
% in Figure \ref{fig:mistral-errors}
did not seem to be strongly impacted by prompt techniques as they did for other LLMs. Note that these differences in the models do not necessarily come from the model size and the number of parameters it was trained on, but rather the underlying architecture it uses (which aligns with findings by Wang et al. \cite{wang2024advanced}).
This implies that when a software company integrates an LLM into its processes and provides employee training, it should develop specific guidelines tailored to the LLM, including recommendations for prompt techniques that align with the model's characteristics e.g., if an LLM returns simple functions in general, so prompt techniques that impact complexity may not be needed.

 % invest in a more powerful LLM or prompt engineering? still a question to be asked.

\textbf{I4: Determining the purpose of the code generation is essential for the use of prompt programming}
Depending on whether the intended use of the LLM is to support human developers or to completely automate code generation, prompt programming has different significance.
% when the main purpose of the LLM is to assist software developers in writing code more efficiently, fixing the code and adapting it to a specific context becomes affordable.
\revised{R2C2}{2}{We manually inspected 40 randomly selected failing functions from different code levels, each of which had passed with at least one other prompt, to understand what caused the failures. We saw that while the use of some prompt techniques has significantly minimized the number of errors and code smells, many of these issues can be easily fixed by human developers, arguably requiring less time and effort than re-prompting the LLM and applying an additional prompt technique. For instance, the absence of a signature in a prompt causes \texttt{TypeErrors} when the LLM misjudges the number of arguments, or misses that the function is a part of a class until the signature with a \texttt{self} parameter is provided. Moreover, prompting the LLM with few-shot examples reduced \texttt{AssertionErrors} mostly because the original prompt lacked clear specifications for edge cases and input/output formats, which could be picked up from the examples and result in passed tests (see Figure \ref{fig:error-heatmap}). 
}
% The majority of such bugs may be more easily fixed by the human rather than re-prompting the LLM with the correct signature.
On the other hand, prompt programming can be more valuable when the purpose is to automate code generation and return correct and maintainable code without the need for human intervention, especially to apply simple modifications or refactoring actions.

    % the bugs that you avoid by providing the signature are easy to fix manually. these dont need to be worried about unless you want to automate the code generation.
    % simple prompts are better not to confuse the LLM and not limit it.

\subsection{Threats to validity}
% validity threats.
\textbf{External validity.} The main threats to external validity in this study are associated with the prompt techniques, the LLMs and the benchmark we utilized.
\minor{R2C1}{2}{There are many possible prompt techniques that can potentially impact code generation, such as self-collaboration \cite{jiang2024self}, AceCoder \cite{li2024acecoder}, or providing the whole class as context. However, we decided to select common prompt techniques that can be practically applied by a typical software developer in most code generation tasks.}
Another important question is whether the use of more powerful LLMs can result in different findings and eliminate the need for prompt programming. We used three current-generation LLMs during the study, including GPT-4o (200B parameters) and Llama3 (70B parameters). Our replication package~\cite{khojah2024replication} also includes results for older LLMs (GPT-3.5, Llama2), showing that the prompt techniques affecting code generation in this study similarly impact older models. 
% These factors highlight potential limitations in the generalizability of our findings across different prompt techniques, LLMs, and real-world code generation tasks.

% moved to construct validity
% Finally, the representativeness of the benchmark (including generation tasks and functions) is an important aspect. While current benchmarks often include functions that are not as complex as real life tasks, we used CoderEval dataset that is based on large open-source projects to minimize this threat.

\textbf{Internal validity.} Regarding internal validity, we acknowledge that LLMs can be sensitive to format or structure of the prompt \cite{sclar2023quantifying}, or even the order of the few-shot examples \cite{lu2022fantastically}. \revised{R1C8}{1}{To address this, we manually refine the prompts and ensure that they have the same level of detail, for example, by removing examples that may be described in the original prompt to not impact the few-shot analysis. We also used a fixed order of the prompt techniques that we believe represents a natural sentence flow, and we ensured to use it consistently across all prompts.}
In addition, the manual creation of few-shot examples for our dataset may have introduced a degree of subjectivity. We therefore involve the first three authors in the process, allowing them to discuss possible examples and select the two most representative ones. 
% \minor{R1C1}{1}{Furthermore, we treat the function signature as a prompt technique rather than a constraint, since it is uncommon in code generation benchmarks.}
\minor{R1C2}{1}{Furthermore, to avoid failures due to incorrect function names, we replace the generated signature with the correct one before testing, though failures due to incorrect parameter names may still occur.}
\revised{R1C3}{1}{Finally, since we used recent LLMs, they may have been trained on the same open-source GitHub code we used. To reduce this risk, we avoided code-focused models like Codex and CodeLlama, which are known to be trained specifically on GitHub data. We also checked for memorization by following the method from Schäfer et al. \cite{schafer2024memorization} using the \textit{maximum similarity} metric. \minor{R1C3}{1}{Similarity was found to produce more meaningful results when identifying memorization compared to other techniques that focus solely on code structure \cite{lemieux2023codemosa}. However, it may still miss other forms of memorization, particularly those involving structural overlap.} We found that for all models, 85\% of the generated functions had a maximum similarity score below 0.4, and none were higher than 0.7. This suggests that the models produced solutions based on understanding the input, not memorizing training data.}

\textbf{Construct validity.} The representativeness of the benchmark (including generation tasks and functions) is an important aspect of construct validity. While current benchmarks often include functions that are not as complex as real-life tasks, we used the CoderEval dataset based on large open-source projects to minimize this threat. However, there remains the question of whether CoderEval fully captures the complexity and diversity of real-world development tasks. 
% Additionally, our selection of practical prompt techniques, defined as those commonly applied by typical software developers, could oversimplify the concept of "practicality" and fail to account for scenarios where more advanced prompt techniques may be necessary.

\textbf{Conclusion validity.} 
% For conclusion validity, we consider our evaluation approach. We focused on three key criteria: similarity, correctness, and quality. While additional criteria, such as efficiency, could have been considered, we argue that the ones we used are sufficient in the context of our experiment to answer our research questions. Moreover, we ensure robustness by using multiple metrics for each of the three selected criteria. 
For conclusion validity, we focused on three key criteria: similarity, correctness, and quality. While others, like efficiency, could be considered, we argue these suffice for our research questions. Robustness is ensured with multiple metrics for each criterion.

\section{Conclusion}

In this study, we have investigated the impact of different prompt techniques on code generation, specifically function synthesis, along three quality dimensions (correctness, similarity to a human-written baseline, and code quality). We studied five prompt techniques, namely few-shot learning, automatic chain-of-thought, providing a persona, providing a signature, and listing packages. We conduct a full factorial analysis of these five factors using CodePromptEval dataset, which we developed based on CoderEval. We studied three current-generation LLMs, namely GPT-4o, Llama3, and Mistral.

Our key lessons learned were that the impact of prompt techniques on correctness, similarity, and quality was not as large as might be expected. Most combinations of prompt techniques do not lead to statistically significant improvements (or regressions) in correctness, quality, or similarity. Providing type information for the function that is to be generated, either explicitly through a signature, or implicitly via few-shot examples, has the most clear positive effect, particularly on correctness. Some prompt techniques have a positive impact on correctness, and others on quality. However, the obvious idea of combining them usually improves neither.

A possible future extension of our research is to evaluate to what extent our findings generalize to other code generation tasks (e.g., line completion, program repair, or the generation of full applications) \minor{R2C1}{2}{and other prompt techniques}. It is plausible that some of the prompt techniques that did not show a meaningful positive impact on correctness in our experiments (e.g., chain-of-thought) turn out to be more relevant if the generation task is more complex. \minor{R2C2}{2}{Additionally, there are other quality metrics, such as performance or energy efficiency, which should be studied in future work --- particularly given that recent work indicates that AI-generated code frequently exhibits performance regressions~\cite{shuangli2024performance}.}

\section*{Acknowledgements}
This work was partially supported by the Wallenberg AI, Autonomous Systems and Software Program (WASP) funded by the Knut and Alice Wallenberg Foundation.
Additionally, LLM executions were enabled by resources provided by the National Academic Infrastructure for Supercomputing in Sweden (NAISS), partially funded by the Swedish Research Council through grant agreement no. 2022-06725.

\bibliographystyle{ieeetr}
\bibliography{bib}

\end{document}